\documentclass[12pt,a4paper]{article}%
\usepackage{graphicx}
\usepackage{amsmath}
\usepackage{makeidx}%
\usepackage{amsfonts}%
\usepackage{amssymb}

\newcommand{\be}{\begin{equation}}
\newcommand{\ee}{\end{equation}}
\newcommand{\beq}{\begin{equation}}
\newcommand{\eeq}{\end{equation}}
\newcommand{\bea}{\begin{eqnarray}}
\newcommand{\eea}{\end{eqnarray}}
\newcommand{\bdm}{\begin{displaymath}}
\newcommand{\edm}{\end{displaymath}}

\begin{document}

\title{{\LARGE \textbf{Supersymmetric models with minimal flavour violation and their
running}}\\[8mm]}
\author{\textbf{Gilberto Colangelo}, \textbf{Emanuel Nikolidakis}\\and \textbf{Christopher Smith}\\[2mm] {\small Institut f\"ur Theoretische Physik, Universit\"at Bern}\\{\small Sidlerstr. 5, 3012 Bern, Switzerland}}
\date{July 4, 2008}
\maketitle

\begin{abstract}
We revisit the formulation of the principle of minimal flavor violation (MFV)
in the minimal supersymmetric extension of the standard model, both at
moderate and large $\tan\beta$, and with or without new CP-violating phases.
We introduce a counting rule which keeps track of the highly hierarchical
structure of the Yukawa matrices. In this manner, we are able to control
systematically which terms can be discarded in the soft SUSY breaking part of
the Lagrangian. We argue that for the implementation of this counting rule, it
is convenient to introduce a new basis of matrices in which both the squark
(and slepton) mass terms as well as the trilinear couplings can be expanded.
We derive the RGE for the MFV parameters and show that the beta functions also
respect the counting rule. For moderate $\tan\beta$, we provide explicit
analytic solutions of these RGE and illustrate their behaviour by analyzing
the neighbourhood (also switching on new phases) of the SPS-1a benchmark
point. We then show that even in the case of large $\tan\beta$, the RGE remain
valid and that the analytic solutions obtained for moderate $\tan\beta$ still
allow us to understand the most important features of the running of the
parameters, as illustrated with the help of the SPS-4 benchmark point.

\pagebreak 

\end{abstract}
\tableofcontents

\thispagestyle{empty}  \setcounter{page}{0}  \clearpage                        


\newpage

\section{Introduction}

In the Standard Model, renormalizability restricts the possible sources of
flavour violations to only one matrix, the Cabibbo-Kobayashi-Maskawa (CKM)
matrix. Its almost diagonal structure (like the values of all the other free
parameters of the model) remains unexplained.

In extensions of the Standard Model, as soon as new degrees of freedom appear,
the possibilities to generate transitions among flavours increase very
rapidly, and may give rise to a richer phenomenology than what the Standard
Model alone would allow. The wealth of recent experimental results produced at
kaon and $B$ factories shows, however, that if new degrees of freedom exist
just above the electroweak symmetry breaking scale, their influence on
low-energy flavour physics is smaller than one would naively expect, below the
current experimental sensitivity. This imposes a nontrivial constraint for
model building, and forces one to impose some sort of protection against
flavour violations.

A convenient way to do this, without excluding completely the possibility to
have new effects in flavour physics, is the principle of minimal flavour
violation (MFV) \cite{Hall:1990ac,D'Ambrosio:2002ex} (see also
Ref.~\cite{Ciuchini:1998xy}). According to this, even in extensions of the
Standard Model, the only source of flavour violations is in the Yukawa
matrices, and since one of them can always be diagonalized, in the CKM matrix.
In its latest, more complete implementation, the principle is formulated as a
symmetry: one starts from the observation that the large global flavour
symmetry group of the Standard Model in the absence of the Yukawa couplings is
saved even in their presence if they are promoted to the status of spurions,
i.e. if they are assigned transformation properties under the flavour group.
The principle of MFV requires that the same symmetry holds even in extensions
of the Standard Model -- the only allowed spurions being the Yukawa matrices.

In this paper, we concentrate on supersymmetric extensions of the Standard
Model with minimal field content and exact R-parity (MSSM) (for a nice
introduction, cf. Ref.~\cite{Martin:1997ns}) and discuss the principle of MFV
within this framework. Our main points are:

\begin{enumerate}
\item We observe that, if one considers the Yukawa matrices as dimension-zero
spurions, and allows any power of them to appear in local operators, imposing
MFV on the MSSM does not restrict the number of free parameters of the model,
but amounts to a mere reparametrization. Still, if one requires that the
coupling constant of the model are of order one, some of them are irrelevant
for the phenomenology and can therefore be dropped.

\item In order to decide systematically which terms are irrelevant and which
ones should be kept, we find it convenient to introduce a counting rule, and
use as expansion parameter $\lambda$, the Cabibbo angle. We use this expansion
parameter to take into account not only the highly hierarchical CKM matrix,
but also the highly hierarchical quark and lepton masses. Each of the MFV
parameters will therefore be assigned an order in $\lambda$ -- once one has
set the accuracy of its calculation to a certain level, $\mathcal{O}%
(\lambda^{n})$, it is immediate to see which MFV parameters should be kept.

\item Since imposing MFV amounts to a mere reparametrization of the generic
MSSM, it is obvious that the principle is renormalization group (RG) invariant
-- what is not guaranteed is that the coefficients will keep their order in
$\lambda$ during the running. We will rewrite the renormalization group
equations (RGE) in MFV form, checking that the beta functions also respect the
same counting rules as the parameters themselves. As such, this is only a
necessary, but not yet sufficient condition to prove that the MFV principle is
truly RGE invariant. We will therefore study the solutions of the RGE
numerically, and show that indeed they respect the counting rules when
evolving from the high to the low scale -- none of the irrelevant parameters
at the high scale may become so large during the RG evolution that it becomes
of phenomenological importance at the low scale. We will also show that the
converse is not true: generic MFV-like boundary conditions at the low scale
will not necessarily evolve to a MFV-like MSSM at the high scale. Therefore,
the assumption that the MFV hypothesis in the MSSM is valid at different
scales makes it even more restrictive at the low scale.

\item We discuss in detail the possible new CP-violating phases allowed by the
MFV hypothesis, and analyze their behaviour under the running. We show that
they tend to vanish at the low scale -- how fast they do that depends on the
initial conditions at the high scale.
\end{enumerate}

While this paper was being completed, a preprint appeared which also analyzes
the behaviour of MFV models under running \cite{Paradisi:2008qh}. There, the
numerical analysis is performed with the help of SOFTSUSY
\cite{Allanach:2001kg}, one of the available codes which allow one to run the
MSSM with a generic flavour structure according to the full RGE to two loops.
In this paper, the authors start with MFV-compatible initial conditions at the
GUT scale, evolve all the parameters down to the electroweak scale, and
project back the model on the MFV parameters, with the help of a fit. Their
analysis is valid for moderate $\tan\beta$, and only for real MFV
coefficients. In this manner, they find out that the MFV parameters have quasi
fixed points at the low scale. We will confirm their finding, and also provide
an analytical explanation for this behaviour. Furher, we will perform the
analysis also for large $\tan\beta$, and in the presence of new CP-violating
phases in the MFV expansions.

\section{Revisiting minimal flavour violation}

\subsection{Definition of minimal flavour violation}

Gauge interactions in the Standard Model are flavour blind. If one sets the
Yukawa matrices to zero, the Standard Model becomes invariant under a large
global symmetry group $G_{F}\sim\lbrack U(3)]^{5}$ \cite{Chivukula:1987py}:
\begin{equation}
G_{F}\equiv G_{q}\otimes G_{\ell}\otimes U(1)_{B}\otimes U(1)_{L}\otimes
U(1)_{Y}\otimes U(1)_{PQ}\otimes U(1)_{E_{R}}\;,
\end{equation}
where
\begin{equation}
G_{q}\equiv SU(3)_{Q_{L}}\otimes SU(3)_{U_{R}}\otimes SU(3)_{D_{R}}\;,\quad
G_{\ell}\equiv SU(3)_{L_{L}}\otimes SU(3)_{E_{R}}\;\;.
\end{equation}
The five $U(1)$ factors have been decomposed in the three which remain a
symmetry even in the presence of Yukawa interactions (related to baryon and
lepton number and hypercharge), and the remaining two. Following
Ref.~\cite{D'Ambrosio:2002ex}, we write these as a phase transformation
affecting $D_{R}$ and $E_{R}$ at the same time, the Peccei-Quinn symmetry of
the two-Higgs doublet model (denoted here by $U(1)_{PQ}$) and one affecting
only $E_{R}$.

The Yukawa matrices break $G_{q}\otimes G_{\ell}\otimes U(1)_{PQ}\otimes
U(1)_{E_{R}}$:
\begin{equation}
\mathcal{L}_{\scriptscriptstyle  Y}=\bar{U}_{R}\,\mathbf{Y}_{u}\,Q_{L}%
H+\bar{D}_{R}\,\mathbf{Y}_{d}\,Q_{L}H_{c}+E_{R}\,\mathbf{Y}_{e}\,L_{L}%
H_{c}+\mathrm{h.c.}\;\;,
\end{equation}
where $H_{c}=i\tau_{2}H^{\ast}$. As observed in Ref.~\cite{D'Ambrosio:2002ex},
the Standard Model remains formally invariant under $G_{q}\otimes G_{\ell}$ in
the presence of the Yukawa matrices if these are promoted to spurion fields
transforming as
\begin{equation}
\mathbf{Y}_{u}\sim(\bar{3},3,1)\;,\quad\mathbf{Y}_{d}\sim(\bar{3}%
,1,3)\quad\mbox{under
}  G_{q}\;,
\end{equation}
and
\begin{equation}
\mathbf{Y}_{e}\sim(\bar{3},3)\quad\mbox{under }  G_{\ell}\;\;.
\end{equation}
The symmetry is broken whenever the Yukawa matrices are frozen at a certain
value -- on the other hand, different forms of the Yukawa matrices which are
related by $G_{q}$ and $G_{\ell}$ transformations are physically equivalent.
In what follows, we will choose the following background values%
\begin{equation}
\mathbf{Y}_{u}=\lambda_{u}V\;,\quad\mathbf{Y}_{d}=\lambda_{d}\;,\quad
\mathbf{Y}_{e}=\lambda_{e}\;, \label{eq:Ybackv}%
\end{equation}
where $\lambda_{u}=\mathrm{diag}(y_{u},y_{c},y_{t})$, $\lambda_{d}%
=\mathrm{diag}(y_{d},y_{s},y_{b})$ and $\lambda_{e}=\mathrm{diag}(y_{e}%
,y_{\mu},y_{\tau})$, and $V$ is the CKM matrix.

An extension of the Standard Model is said to respect MFV if it is symmetric
under $G_{q}\otimes G_{\ell}$ in the presence of Yukawa spurions. While new
matter fields in such an extension are of course allowed, one is not supposed
to introduce new spurion fields beyond the Yukawa matrices.

\subsection{MFV in the MSSM: a reparametrization of the soft SUSY breaking terms}

In a supersymmetric extension of the Standard Model, the superpotential
automatically satisfy the MFV principle. On the other hand, MFV strongly
constrains the soft supersymmetry breaking terms. We will illustrate this
statement by considering the mass term for the left-handed squarks
\begin{equation}
\mathcal{L}_{\mathbf{m}_{Q}^{2}}\equiv-\tilde{Q}^{\dagger}\mathbf{m}_{Q}%
^{2}\cdot\tilde{Q}\;\;,
\end{equation}
and showing first that MFV amounts in principle to a reparametrization of a
generic hermitian $3\times3$ matrix, and later that if one excludes the
possibility of having enormous coupling constants, MFV is indeed quite
constraining. MFV requires this term to become formally invariant under
$G_{q}$, hence it must transform like $(8,1,1)$. Since we are not allowed to
introduce new spurions, we have to obtain this transformation property with
the help of the Yukawa matrices, as with terms like $\mathbf{Y}_{u}^{\dagger
}\mathbf{Y}_{u}$ or $\mathbf{Y}_{d}^{\dagger}\mathbf{Y}_{d}$. Moreover, one
can construct invariants under $G_{q}$ also with the help of $\epsilon
$-tensors. Such terms have recently been systematically studied in
Ref.~\cite{Nikolidakis:2007fc}, and permit to extend the MFV principle to the
R-parity violating interactions. Interestingly, MFV alone is then sufficient
to prevent the proton from decaying too rapidly -- a fact which lends
additional support to the validity of the MFV hypothesis at low energy.

For the R-parity conserving sector of the MSSM, on which we concentrate in the
present paper, these $\epsilon$-tensor terms have been shown to be very
suppressed \cite{Nikolidakis:2007fc}. Hence, ignoring them altogether, MFV
permits to write $\mathbf{m}_{Q}^{2}$ as the infinite sum:
\begin{equation}\label{jwhk}
\mathbf{m}_{Q}^{2}=a_{1}\mathbf{1}+b_{1}\mathbf{Y}_{u}^{\dagger}\mathbf{Y}%
_{u}+b_{2}\mathbf{Y}_{d}^{\dagger}\mathbf{Y}_{d}+c_{1}\mathbf{Y}_{u}^{\dagger
}\mathbf{Y}_{u}\mathbf{Y}_{u}^{\dagger}\mathbf{Y}_{u}+c_{2}\mathbf{Y}%
_{d}^{\dagger}\mathbf{Y}_{d}\mathbf{Y}_{d}^{\dagger}\mathbf{Y}_{d}+\ldots
\end{equation}
Actually, the infinite sum collapses on its first few terms: the Yukawa
couplings are $3\times3$ matrices, hence they respect the corresponding
Cayley-Hamilton identities. The hermitian matrix 
$\mathbf{m}_{Q}^{2}$ does not contain more than nine independent real 
parameters and it can be shown that the sum in Eq.~(\ref{jwhk}) spans 
the space of hermitian matrices \cite{nikphd}.

The Cayley-Hamilton identities read for $3\times3$ matrices 
\begin{equation}
\mathbf{X}^{3}-\left\langle \mathbf{X}\right\rangle \mathbf{X}^{2}+\frac{1}%
{2}\mathbf{X}\left(  \left\langle \mathbf{X}\right\rangle ^{2}-\left\langle
\mathbf{X}^{2}\right\rangle \right)  -\det\mathbf{X}=0\;, \label{eq:CH1}%
\end{equation}
and can be rewritten in terms of traces only if the determinant is expressed
as
\begin{equation}
\det\mathbf{X}=\frac{1}{3}\left\langle \mathbf{X}^{3}\right\rangle
-\frac{1}{2}\left\langle \mathbf{X}\right\rangle \left\langle \mathbf{X}%
^{2}\right\rangle +\frac{1}{6}\left\langle \mathbf{X}\right\rangle ^{3}\;\;.
\end{equation}
In other words, all powers of three or more of a combination of Yukawa
matrices, $\mathbf{X}^{n>2}$, can be eliminated in terms of only
$\mathbf{X}^{2},\mathbf{X},\mathbf{1}$, with coefficients involving the trace
of $\mathbf{X}^{2}$ and $\mathbf{X}$. Further, identities involving two (or
more) different combinations, $\mathbf{A}$ and $\mathbf{B}$ say, can be found
by substituting $\mathbf{X}=a\mathbf{A}+b\mathbf{B}$ in Eq.~(\ref{eq:CH1}) and
extracting a given power of $a$ and $b$. For example, a relevant identity is%
\begin{align}\label{xy2}
&  \!\!\!\!\!\!\mathbf{A}^{2}\mathbf{B}+\mathbf{ABA}+\mathbf{BA}%
^{2}=\mathbf{A}^{2}\left\langle \mathbf{B}\right\rangle +\left(
\mathbf{AB}+\mathbf{BA}\right)  \left\langle \mathbf{A}\right\rangle
+\mathbf{A}\left(  \left\langle \mathbf{AB}\right\rangle -\left\langle
\mathbf{A}\right\rangle \left\langle \mathbf{B}\right\rangle \right)
\nonumber\\
&  \!\!\!\!\!\!+\frac{1}{2}\mathbf{B}\left(  \left\langle \mathbf{A}%
^{2}\right\rangle -\left\langle \mathbf{A}\right\rangle ^{2}\right)
+\frac{1}{2}\left\langle \mathbf{B}\right\rangle \left(  \left\langle
\mathbf{A}\right\rangle ^{2}-\left\langle \mathbf{A}^{2}\right\rangle \right)
+\left\langle \mathbf{A}^{2}\mathbf{B}\right\rangle -\left\langle
\mathbf{A}\right\rangle \left\langle \mathbf{AB}\right\rangle \;,
\end{align}
with $\mathbf{A}=\mathbf{Y}_{u}^{\dagger}\mathbf{Y}_{u}$ and $\mathbf{B}%
=\mathbf{Y}_{d}^{\dagger}\mathbf{Y}_{d}$.

Taking into account these identities, the most general expression for
$\mathbf{m}_{Q}^{2}$ respecting MFV becomes
\begin{align}
\mathbf{m}_{Q}^{2}  &  =z_{1}\mathbf{1}+z_{2}\mathbf{Y}_{u}^{\dagger
}\mathbf{Y}_{u}+z_{3}\mathbf{Y}_{d}^{\dagger}\mathbf{Y}_{d}+z_{4}%
(\mathbf{Y}_{u}^{\dagger}\mathbf{Y}_{u})^{2}+z_{5}(\mathbf{Y}_{d}^{\dagger
}\mathbf{Y}_{d})^{2}\nonumber\\
&  +z_6 \left(\mathbf{Y}_{d}^{\dagger}\mathbf{Y}_{d}\mathbf{Y}_{u}^{\dagger
}\mathbf{Y}_{u}+\mathrm{h.c.} \right)+z_{7}\mathbf{Y}_{u}^{\dagger}\mathbf{Y}%
_{u}\mathbf{Y}_{d}^{\dagger}\mathbf{Y}_{d}\mathbf{Y}_{u}^{\dagger}%
\mathbf{Y}_{u}\nonumber\\
&  +z_{8}\mathbf{Y}_{d}^{\dagger}\mathbf{Y}_{d}\mathbf{Y}_{u}^{\dagger
}\mathbf{Y}_{u}\mathbf{Y}_{d}^{\dagger}\mathbf{Y}_{d}+z_{9}\left(
  (\mathbf{Y}_{u}^{\dagger}\mathbf{Y}_{u})^{2}(\mathbf{Y}_{d}^{\dagger}
 \mathbf{Y}_{d} )^{2}+\mathrm{h.c.}\right) \nonumber\\
&+iz_{10}(\mathbf{Y}_{d}^{\dagger}\mathbf{Y}_{d}\mathbf{Y}_{u}^{\dagger
}\mathbf{Y}_{u}-\mathrm{h.c.} )+iz_{11} \left(
(\mathbf{Y}_{u}^{\dagger}\mathbf{Y}_{u})^{2}
\mathbf{Y}_{d}^{\dagger}\mathbf{Y}_{d}- \mathrm{h.c.}\right ) \nonumber \\
&+iz_{12} \left(
(\mathbf{Y}_{d}^{\dagger}\mathbf{Y}_{d})^{2}
\mathbf{Y}_{u}^{\dagger}\mathbf{Y}_{u}- \mathrm{h.c.}\right ) 
+iz_{13}\left(
  (\mathbf{Y}_{u}^{\dagger}\mathbf{Y}_{u})^{2}(\mathbf{Y}_{d}^{\dagger}
 \mathbf{Y}_{d} )^{2}-\mathrm{h.c.}\right) \nonumber\\
&+i z_{14}\left( \mathbf{Y}_{u}^{\dagger}\mathbf{Y}_{u}
  \mathbf{Y}_{d}^{\dagger} \mathbf{Y}_{d}
  (\mathbf{Y}_{u}^{\dagger}\mathbf{Y}_{u})^{2}-\mathrm{h.c.}\right) 
+i z_{15}\left( \mathbf{Y}_{d}^{\dagger}\mathbf{Y}_{d}
  \mathbf{Y}_{u}^{\dagger} \mathbf{Y}_{u}
  (\mathbf{Y}_{d}^{\dagger}\mathbf{Y}_{d})^{2}-\mathrm{h.c.}\right)  \nonumber\\
&+i z_{16}\left( \mathbf{Y}_{u}^{\dagger}\mathbf{Y}_{u}
  (\mathbf{Y}_{d}^{\dagger} \mathbf{Y}_{d})^{2}
  (\mathbf{Y}_{u}^{\dagger}\mathbf{Y}_{u})^{2}-\mathrm{h.c.}\right) 
+i z_{17}\left( \mathbf{Y}_{d}^{\dagger}\mathbf{Y}_{d}
  (\mathbf{Y}_{u}^{\dagger} \mathbf{Y}_{u})^{2}
  (\mathbf{Y}_{d}^{\dagger}\mathbf{Y}_{d})^{2}-\mathrm{h.c.}\right) 
\;\;, \label{eq:mq2g}%
\end{align}
with the $z_{i}$ being real parameters. A generic $3\times3$ hermitian matrix can
be described by nine real constants -- in Eq.~(\ref{eq:mq2g}),
$\mathbf{m}_{Q}^{2}$ is expressed in terms of seventeen real constants, so
that eight of them must be linearly dependent. 
Even if we do not specify the
linear relations which allow one to eliminate eight of these
constants\footnote{Note that such a linear relation can be nontrivial and
  may involve large coefficients. We stress that the Cayley-Hamilton
  identities do not involve any large numerical coefficients, and so do not
  upset the assumption that the MFV coefficients are of order one.}, it is
clear that MFV amounts to a mere reparametrization of the soft
SUSY-breaking terms of the MSSM, since the original expansion in 
Eq.~(\ref{jwhk}) contains a basis and the Cayley-Hamilton relations 
are exact. A similar argument can be used also for all other terms.

What is special about the MFV parametrization is that if all the $z_{i}$'s are
of the same order of magnitude, the structure of $\mathbf{m}_{Q}^{2}$ is
highly non-generic. Conversely, if one writes down a generic $\mathbf{m}%
_{Q}^{2}$ matrix and projects it on the MFV basis, the coefficients $z_{i}$ so
obtained will typically span many orders of magnitude. We define extensions of
the Standard Model respecting MFV by the additional requirement that the
coefficients appearing in front of the various MFV terms are of the same order
of magnitude.

\subsection{Counting rules and a new basis}

If one takes all the $z_{i}$ coefficients to be of the same order of
magnitude, several of the terms in Eq.~(\ref{eq:mq2g}) (and in the analogous
ones for the other soft SUSY-breaking terms) can be disposed of. In this
subsection, we will discuss how to do this in a systematic way, and will argue
that it is more convenient to change basis in order to work with MFV. For
example, taking into account the actual values of the Yukawa coefficients of
the up, charm and top quarks, we conclude that the two matrices
\begin{equation}
\left(  \mathbf{Y}_{u}^{\dagger}\mathbf{Y}_{u}\right)  ^{2}=V^{\dagger}%
\lambda_{u}^{4}V\;,\qquad\mathbf{Y}_{u}^{\dagger}\mathbf{Y}_{u}=V^{\dagger
}\lambda_{u}^{2}V
\end{equation}
are proportional to each other up to a correction of relative order
$\mathcal{O}(y_{c}^{2}/y_{t}^{2})$,
\begin{equation}
\left(  \mathbf{Y}_{u}^{\dagger}\mathbf{Y}_{u}\right)  ^{2}-y_{t}%
^{2}\mathbf{Y}_{u}^{\dagger}\mathbf{Y}_{u}=y_{t}^{2}y_{c}^{2}V_{2i}^{\ast
}V_{2j}+\mathcal{O}(y_{c}^{4})\;, \label{eq:yuyu}%
\end{equation}
which is usually neglected since $y_{c}^{2}\ll1$. A similar argument can be
applied to $\mathbf{Y}_{d}^{\dagger}\mathbf{Y}_{d}$, since $y_{s}^{2}\ll1$
also. We will therefore never include any power of $\mathbf{Y}_{u}^{\dagger
}\mathbf{Y}_{u}$ or $\mathbf{Y}_{d}^{\dagger}\mathbf{Y}_{d}$ in our analysis,
and keep the latter to allow $\tan\beta$ to be large (remember that the two
Higgs doublets of the MSSM separately give mass to the up and down-quarks:
$v_{u}\lambda_{u}=diag(m_{u},m_{c},m_{t})$ and $v_{d}\lambda_{d}%
=diag(m_{d},m_{s},m_{b})$, with $v_{u,d}$ the two Higgs vacuum expectation
values, and $\tan\beta\equiv v_{u}/v_{d}$). With this approximation, the MFV
version of the soft SUSY-breaking terms reads\footnote{The numbering of the
coefficients follows the choice of Ref.~\cite{D'Ambrosio:2002ex} whenever
possible. The term $b_{4}$ present in that paper has to be equal to
$b_{3}^{\ast}$ in order to satisfy the hermiticity of the matrix
$\mathbf{m}_{Q}^{2}$.}
\begin{align}
\mathbf{m}_{Q}^{2}  &  =m_{0}^{2}\left[  a_{1}+b_{1}\mathbf{Y}_{u}^{\dagger
}\mathbf{Y}_{u}+b_{2}\mathbf{Y}_{d}^{\dagger}\mathbf{Y}_{d}+(b_{3}%
\mathbf{Y}_{d}^{\dagger}\mathbf{Y}_{d}\mathbf{Y}_{u}^{\dagger}\mathbf{Y}%
_{u}+\mathrm{h.c.})\right]  \;,\nonumber\\
\mathbf{m}_{U}^{2}  &  =m_{0}^{2}\left[  a_{2}+\mathbf{Y}_{u}\!\left(
b_{5}+c_{1}\mathbf{Y}_{u}^{\dagger}\mathbf{Y}_{u}+c_{2}\mathbf{Y}_{d}%
^{\dagger}\mathbf{Y}_{d}+(c_{3}\mathbf{Y}_{d}^{\dagger}\mathbf{Y}%
_{d}\mathbf{Y}_{u}^{\dagger}\mathbf{Y}_{u}+\mathrm{h.c.})\right)
\!\mathbf{Y}_{u}^{\dagger}\right]  \;,\nonumber\\
\mathbf{m}_{D}^{2}  &  =m_{0}^{2}\left[  a_{3}+\mathbf{Y}_{d}\!\left(
b_{6}+c_{4}\mathbf{Y}_{u}^{\dagger}\mathbf{Y}_{u}+c_{5}\mathbf{Y}_{d}%
^{\dagger}\mathbf{Y}_{d}+(c_{6}\mathbf{Y}_{d}^{\dagger}\mathbf{Y}%
_{d}\mathbf{Y}_{u}^{\dagger}\mathbf{Y}_{u}+\mathrm{h.c.})\right)
\!\mathbf{Y}_{d}^{\dagger}\right]  \;,\nonumber\\
\mathbf{A}^{U}  &  =A_{0}\mathbf{Y}_{u}\!\left(  a_{4}+b_{7}\mathbf{Y}%
_{d}^{\dagger}\mathbf{Y}_{d}+c_{7}\mathbf{Y}_{u}^{\dagger}\mathbf{Y}_{u}%
+c_{8}\mathbf{Y}_{d}^{\dagger}\mathbf{Y}_{d}\mathbf{Y}_{u}^{\dagger}%
\mathbf{Y}_{u}+c_{9}\mathbf{Y}_{u}^{\dagger}\mathbf{Y}_{u}\mathbf{Y}%
_{d}^{\dagger}\mathbf{Y}_{d}\right)  \;,\nonumber\\
\mathbf{A}^{D}  &  =A_{0}\mathbf{Y}_{d}\!\left(  a_{5}+b_{8}\mathbf{Y}%
_{u}^{\dagger}\mathbf{Y}_{u}+c_{10}\mathbf{Y}_{d}^{\dagger}\mathbf{Y}%
_{d}+c_{11}\mathbf{Y}_{d}^{\dagger}\mathbf{Y}_{d}\mathbf{Y}_{u}^{\dagger
}\mathbf{Y}_{u}+c_{12}\mathbf{Y}_{u}^{\dagger}\mathbf{Y}_{u}\mathbf{Y}%
_{d}^{\dagger}\mathbf{Y}_{d}\right)  \;. \label{eq:inter1}%
\end{align}

Until now, we have used only the fact that the Yukawa matrices are highly
hierarchical along their diagonal. They are, however, also hierarchical in
their off-diagonal structure, and taking this into account leads to further
simplifications. In order to do this in a systematic way, we use as expansion
parameter the Cabibbo angle $\lambda=0.23$, which appears in the Wolfenstein
parametrization of the CKM matrix as follows (to leading order in $\lambda$ --
in subsequent calculations we will always include higher orders also)
\begin{equation}
V\approx\left(
\begin{array}
[c]{ccc}%
1 & \lambda & A\lambda^{3}\left(  \rho-i\eta\right) \\
-\lambda & 1 & A\lambda^{2}\\
A\lambda^{3}\left(  1-\rho-i\eta\right)  & -A\lambda^{2} & 1
\end{array}
\right)  \;, \label{eq:CKMW}%
\end{equation}
We then adopt the following counting conventions for the quark mass ratios at
the electroweak scale $\mu=M_{Z}$:
\begin{align}
&  \frac{m_{u}}{m_{t}}\sim\mathcal{O}(\lambda^{7}),\;\;\frac{m_{c}}{m_{t}}%
\sim\mathcal{O}(\lambda^{4}),\;\;y_{t}\sim\mathcal{O}(1)\;,\label{eq:counting}%
\\
&  \frac{m_{d}}{m_{t}}\sim\mathcal{O}(\lambda^{7}),\;\;\frac{m_{s}}{m_{t}}%
\sim\mathcal{O}(\lambda^{5}),\;\;\frac{m_{b}}{m_{t}}\sim\mathcal{O}%
(\lambda^{3})\; .\nonumber
\end{align}

For example, the difference in Eq.~(\ref{eq:yuyu}),
\begin{equation}
\mathbf{Y}_{u}^{\dagger}\mathbf{Y}_{u}\mathbf{Y}_{u}^{\dagger}\mathbf{Y}%
_{u}-y_{t}^{2}\mathbf{Y}_{u}^{\dagger}\mathbf{Y}_{u}=\left(
\begin{array}
[c]{ccc}%
\mathcal{O}\left(  \lambda^{8}\right)  & \mathcal{O}\left(  \lambda^{7}\right)
& \mathcal{O}\left(  \lambda^{9}\right) \\
\mathcal{O}\left(  \lambda^{7}\right)  & \mathcal{O}\left(  \lambda^{6}\right)
& \mathcal{O}\left(  \lambda^{8}\right) \\
\mathcal{O}\left(  \lambda^{9}\right)  & \mathcal{O}\left(  \lambda^{8}\right)
& \mathcal{O}\left(  \lambda^{10}\right)
\end{array}
\right)  \;,
\end{equation}
represents a correction of at least $\mathcal{O}(\lambda^{2})$ (in every
entry) to each of the two terms, cf.
\begin{equation}
\mathbf{Y}_{u}^{\dagger}\mathbf{Y}_{u}=\left(
\begin{array}
[c]{ccc}%
\mathcal{O}\left(  \lambda^{6}\right)  & \mathcal{O}\left(  \lambda^{5}\right)
& \mathcal{O}\left(  \lambda^{3}\right) \\
\mathcal{O}\left(  \lambda^{5}\right)  & \mathcal{O}\left(  \lambda^{4}\right)
& \mathcal{O}\left(  \lambda^{2}\right) \\
\mathcal{O}\left(  \lambda^{3}\right)  & \mathcal{O}\left(  \lambda^{2}\right)
& \mathcal{O}\left(  1\right)
\end{array}
\right)  \;\;.
\end{equation}
This shows that while MFV can indeed be viewed as a reparametrization, cf.
Eq.~(\ref{eq:mq2g}), it is on the other hand a very special one, because
the basis in the linear space of $3\times3$ hermitian matrices on which
$m_{Q}%
^{2}$ is projected is almost a singular one. Several of the basis vectors
are almost parallel to each other, and their difference is tiny in
comparison to either of the vectors. Since our aim here is to link the MFV
concept to a counting rule, and define clearly which terms should be kept
and which should be neglected, it is a lot more convenient to use a basis
of vectors which are as little \footnote{Strictly speaking one could - in
  contrast to what we will do - consider a fully orthogonal basis of
  matrices (like $e_{ij}^{kl}=\delta_i^k\delta_j^l$). In this way, however,
  one would lose track of the fact that the physically relevant degrees
  of freedom of the Yukawa matrices are contained in two diagonal and one
  unitary matrix.}  aligned to
each other as possible. In this way, contributions which are small and may
(or may not) be neglected will not have to be searched for in small
differences between similar contributions, but will be clearly identified
and separated from the rest. In order to illustrate this concept, we come
back to Eq.~(\ref{eq:yuyu}) and observe that the large piece in both
$\mathbf{Y}_{u}^{\dagger}\mathbf{Y}_{u}$ and its square is proportional to
the matrix $V_{3i}^{\ast}V_{3j}$, whereas the small one, as indicated in
Eq.~(\ref{eq:yuyu}) is proportional to $V_{2i}^{\ast}V_{2j}$. So, instead
of using $\mathbf{Y}_{u}^{\dagger}\mathbf{Y}_{u}$ and
$(\mathbf{Y}_{u}^{\dagger }\mathbf{Y}_{u})^{2}$ as basis vectors, and allow
both to have coefficients of order one, we find it more convenient to use
$V_{3i}^{\ast}V_{3j}$ and $V_{2i}^{\ast}V_{2j}$, and say that the first can
have a coefficient of order one, but the latter should have it of order
$y_{c}^{2}\sim\lambda^{8}$, and (depending on where one sets its accuracy)
can therefore be neglected.

If we follow the same logic for $\mathbf{Y}_{d}^{\dagger}\mathbf{Y}_{d}$ and
its square, we find that in this case we should rather use as basis vectors
the matrices $\delta_{i3}\delta_{j3}$ and $\delta_{i2}\delta_{j2}$, and that
the coefficients of these terms should be of order $y_{b}^{2}$ and $y_{s}^{2}$
respectively (how to translate this into a power of $\lambda$ depends on
$\tan\beta$). Taking into account all the possible structures which can
emerge, and which can all be multiplied with each other, we are led to
consider a set of sixteen matrices, which form a closed algebra under
multiplication:
\begin{equation}%
\begin{array}
[c]{llll}%
X_{1}=\delta_{3i}\delta_{3j} & X_{5}=\delta_{3i}V_{3j} & X_{9}=V_{3i}^{\ast
}\delta_{3j} & X_{13}=V_{3i}^{\ast}V_{3j}\\
X_{2}=\delta_{2i}\delta_{2j} & X_{6}=\delta_{2i}V_{2j} & X_{10}=V_{2i}^{\ast
}\delta_{2j} & X_{14}=V_{2i}^{\ast}V_{2j}\\
X_{3}=\delta_{3i}\delta_{2j} & X_{7}=\delta_{3i}V_{2j} & X_{11}=V_{3i}^{\ast
}\delta_{2j} & X_{15}=V_{3i}^{\ast}V_{2j}\\
X_{4}=\delta_{2i}\delta_{3j} & X_{8}=\delta_{2i}V_{3j} & X_{12}=V_{2i}^{\ast
}\delta_{3j} & X_{16}=V_{2i}^{\ast}V_{3j}%
\end{array}
\label{eq:Mbasis}%
\end{equation}
Notice that all these matrices have at least one entry (almost) equal to one,
so that they can all be counted as of order one.

If we now write Eq.~(\ref{eq:inter1}) (or even start from the earlier stage
given by Eq.~(\ref{eq:mq2g})) in this basis and assume that all $a_{i}$,
$b_{i}$ and $c_{i}$ coefficients are of order one, we can immediately read out
the size of the coefficients in front of the $X_{i}$ matrices and decide which
one to keep and which not. In particular, if we drop terms of order
$\lambda^{6}\sim10^{-4}$ and higher, we can reduce the soft SUSY-breaking
terms to the following
\begin{align}
\mathbf{m}_{Q}^{2}  &  =\tilde{a}_{1}+x_{1}X_{13}+y_{1}X_{1}+y_{2}X_{5}%
+y_{2}^{\ast}X_{9}\;,\nonumber\\
\mathbf{m}_{U}^{2}  &  =\tilde{a}_{2}+x_{2}X_{1}\;,\nonumber\\
\mathbf{m}_{D}^{2}  &  =\tilde{a}_{3}+y_{3}X_{1}+w_{1}X_{3}+w_{1}^{\ast}%
X_{4}\;,\nonumber\\
\mathbf{A}^{U}  &  =\tilde{a}_{4}X_{5}+y_{4}X_{1}+w_{2}X_{6}\;,\nonumber\\
\mathbf{A}^{D}  &  =\tilde{a}_{5}X_{1}+y_{5}X_{5}+w_{3}X_{2}+w_{4}X_{4}\;,
\label{eq:xMFV}%
\end{align}
where, for simplicity, we have absorbed the overall scales $m_{0}^{2}$ and
$A_{0}$ into the coefficients. Since the matrices $X_{i}$ are all of order
one, the coefficients now carry the order in $\lambda$, and we have reflected
this in the symbols which identify them. Relative to the leading terms, the
two $x_{i}$'s are of order one, $\tilde{a}_{5}$ (which now incorporates
$y_{b}$ -- while $\tilde{a}_{4}$ incorporates $y_{t}$), as well as the $y_{i}%
$'s can become of order one if $\tan\beta\sim\lambda^{-3}$, and the $w_{i}$'s
are suppressed by at least two powers of $\lambda$, even when $\tan\beta
\sim\lambda^{-3}$. More specifically:%
\begin{gather}
\frac{\tilde{a}_{5}}{\tilde{a}_{4}}\sim\frac{y_{5}}{\tilde{a}_{4}}%
\sim\mathcal{O}(\lambda^{3}t_{\beta})\;,\;\;\;\;\;\;\;\;\frac{y_{1,2}%
}{\tilde{a}_{1}}\sim\frac{y_{3}}{\tilde{a}_{3}}\sim\frac{y_{4}}{\tilde{a}_{4}%
}\sim\mathcal{O}(\lambda^{6}t_{\beta}^{2})\;,\nonumber\\
\frac{w_{1}}{\tilde{a}_{3}}\sim\mathcal{O}(\lambda^{10}t_{\beta}%
^{2})\;,\;\;\;\frac{w_{2}}{\tilde{a}_{3}}\sim\mathcal{O}(\lambda
^{4})\;,\;\;\;\frac{w_{3}}{\tilde{a}_{5}}\sim\mathcal{O}(\lambda
^{2})\;,\;\;\;\frac{w_{4}}{\tilde{a}_{5}}\sim\mathcal{O}(\lambda^{4})\;.
\label{xilambda}%
\end{gather}
We stress that this representation remains valid even in the case of large
$\tan\beta$, up to $\tan\beta\sim\mathcal{O}(m_{t}/m_{b})$ -- on the other
hand, if the latter is not large, all the $y_{1-4}$, $w_{1}$ and $w_{4}$ can
be dropped. Also, $a_{1-3}$, $x_{1}$, $x_{2}$, $y_{1}$ and $y_{3}$ must be
real since $\mathbf{m}_{Q,U,D}^{2}$ are hermitian, while all the others can be
complex. There are thus a priori nine new CP-violating phases entering in the
squark soft-breaking terms, though only those of $\tilde{a}_{4}$ and (less so)
$\tilde{a}_{5}$ and $y_{5}$ are unsuppressed for moderate $\tan\beta$.

Looking at the representation (\ref{eq:xMFV}), one may observe that, in
contrast to the original formulation of Ref.~\cite{D'Ambrosio:2002ex}, the
symmetry principle which is behind MFV is not immediately visible anymore.
Actually, the essential concept that the only source of flavour violation is
the CKM matrix is still manifest in the $X_{i}$ basis, and moreover:

\begin{enumerate}
\item The symmetry principle which forms the basis of MFV is used once and for
all, and leads to the basis shown, \emph{e.g.} in Eq.~(\ref{eq:mq2g}). The
relation between the standard MFV and the $X_{i}$ basis can also be derived
once and for all; it is given explicitly in Appendix~\ref{AppendixBasis} in
terms of the running Yukawa couplings, to be defined in Sect.~\ref{sec:RGE}.

\item To make the most of the MFV principle, one still has to drop suppressed
terms in the most general representation provided by MFV in
Eq.~(\ref{eq:inter1}). As we have discussed above, it is easier and more
transparent to do so in the $X_{i}$ basis.

\item The $x_{i}$ coefficients\footnote{In what follows we will often use the
expression ``the $x_{i}$ coefficients'' or ``the $x_{i}$'s'' to mean all
$x_{i}$, $y_{i}$ and $w_{i}$ coefficients. In order not to generate confusion,
we will say explicitly when we mean specifically $x_{1}$ and $x_{2}$.} have a
direct, simple relation to the mass insertions \cite{Hall:1985dx}, which are
useful for phenomenological studies of flavour violations, as we will discuss
in detail in the next section.

\item The scalings of the $x_{i}$ coefficients, as well as the decompositions
(\ref{eq:xMFV}), are stable under the RGE's, as will be discussed in details
in Sect.~\ref{sec:RGE}. Further, the formulation of the RGE's for the $x_{i}$
coefficients is particularly convenient, especially because one can
immediately see the order of the different terms appearing in their beta functions.
\end{enumerate}

\subsection{MFV mass-insertions and their impact on phenomenology}

\label{sec:MassI} In many phenomenological applications, it is convenient not
to assume anything about the structure of the soft SUSY breaking terms other
than the established experimental fact that, if present, they are almost
diagonal. One then writes them as diagonal matrices plus small off-diagonal
corrections and, when calculating observables, expands them in the
off-diagonal terms -- called mass insertions \cite{Hall:1985dx}. Of course, in
most practical applications, the exact diagonalization of the squark mass
matrices is performed (see e.g. Refs.~\cite{MFVpheno1,MFVpheno2} for studies
in the MSSM with MFV). But even if the use of mass insertions is thereby
circumvented, they remain a very convenient tool to organize and identify
possible sources of flavour violation. Indeed, the rich experimental
information on flavour violations has been translated in bounds on the mass
insertions \cite{Gabbiani:1996hi,Ciuchini:2007ha}, and it is therefore useful
to provide a relation between the MFV parameters and the latter.

The LL and RR mass-insertions, defined as%
\begin{equation}
(\delta_{LL})^{IJ}=\frac{(\mathbf{m}_{Q}^{2})^{IJ}}{|(\mathbf{m}_{Q}^{2}%
)^{II}|^{1/2}|(\mathbf{m}_{Q}^{2})^{JJ}|^{1/2}}\;,\;\;(\delta_{RR}^{F}%
)^{IJ}=\frac{(\mathbf{m}_{F}^{2})^{IJ}}{|(\mathbf{m}_{F}^{2})^{II}%
|^{1/2}|(\mathbf{m}_{F}^{2})^{JJ}|^{1/2}}\;,
\end{equation}
with $F=U,D$, are in terms of the MFV coefficients%
\begin{align}
(\delta_{LL})^{23}  &  =V_{tb}V_{ts}^{\ast}\frac{x_{1}+y_{2}^{\ast}%
}{|\tilde{a}_{1}|^{1/2}|\tilde{a}_{1}+x_{1}+y_{1}+2\operatorname{Re}%
y_{2}|^{1/2}}=\frac{V_{tb}V_{ts}^{\ast}}{V_{tb}V_{td}^{\ast}}(\delta
_{LL})^{13}\;,\;\;\\
(\delta_{LL})^{12}  &  =V_{ts}V_{td}^{\ast}\frac{x_{1}}{|\tilde{a}_{1}%
|}\;,\;\;(\delta_{LL})^{IJ}=(\delta_{LL})^{JI\ast}\;, \label{MIA1}%
\end{align}
up to completely negligible corrections of relative order $\lambda^{4}$ or
higher. In our approximation, all $\delta_{RR}^{U}=0$, while only one
$\delta_{RR}^{D}$ is non-zero,%
\begin{equation}
(\delta_{RR}^{D})^{23}=\frac{w_{1}^{\ast}}{|\tilde{a}_{3}|^{1/2}%
|\tilde{a}_{3}+y_{3}|^{1/2}}=(\delta_{RR}^{D})^{32\ast}\;, \label{MIA2}%
\end{equation}
but is nevertheless extremely suppressed since $w_{1}\sim\mathcal{O}%
(\lambda^{10}t_{\beta}^{2})$. The $SU(2)_{L}$-breaking RL mass-insertions are
defined as%
\begin{equation}
(\delta_{RL}^{U})^{IJ}=\frac{v_{u}(\mathbf{A}^{U})^{IJ}}{|(\mathbf{m}_{U}%
^{2})^{II}|^{1/2}|(\mathbf{m}_{Q}^{2})^{JJ}|^{1/2}}\;,\;\;(\delta_{RL}%
^{D})^{IJ}=\frac{v_{d}(\mathbf{A}^{D})^{IJ}}{|(\mathbf{m}_{D}^{2})^{II}%
|^{1/2}|(\mathbf{m}_{Q}^{2})^{JJ}|^{1/2}}\;,
\end{equation}
and are%
\begin{align}
(\delta_{RL}^{U})^{32}  &  =V_{ts}\frac{v_{u}\tilde{a}_{4}}{|\tilde{a}_{1}%
|^{1/2}|\tilde{a}_{2}+x_{2}|^{1/2}}=\frac{V_{ts}}{V_{td}}(\delta_{RL}%
^{U})^{31}\;,\nonumber\\
(\delta_{RL}^{U})^{23}  &  =V_{cb}\frac{v_{u}w_{2}}{|\tilde{a}_{2}%
|^{1/2}|\tilde{a}_{1}+x_{1}+y_{1}+2\operatorname{Re}y_{2}|^{1/2}%
}\;,\nonumber\\
(\delta_{RL}^{U})^{21}  &  =V_{cd}\frac{v_{u}w_{2}}{|\tilde{a}_{1}%
|^{1/2}|\tilde{a}_{2}|^{1/2}}\;,\nonumber\\
(\delta_{RL}^{D})^{32}  &  =V_{ts}\frac{v_{d}y_{5}}{|\tilde{a}_{1}%
|^{1/2}|\tilde{a}_{3}+y_{3}|^{1/2}}=\frac{V_{ts}}{V_{td}}(\delta_{RL}%
^{D})^{31}\;,\nonumber\\
(\delta_{RL}^{D})^{23}  &  =\frac{v_{d}w_{4}}{|\tilde{a}_{3}|^{1/2}%
|\tilde{a}_{1}+x_{1}+y_{1}+2\operatorname{Re}y_{2}|^{1/2}}\;, \label{MIA3}%
\end{align}
while $(\delta_{RL}^{U})^{12}=(\delta_{RL}^{U})^{13}=0$ as well as
$(\delta_{RL}^{D})^{13}=(\delta_{RL}^{D})^{21}=(\delta_{RL}^{D})^{12}=0$ under
our approximation of neglecting anything of order $\lambda^{6}$ or higher. The
forms of the various mass-insertions show that indeed, the CKM matrix elements
still tune all the flavour transitions. In addition, the $X_{i}$ basis permits
to immediately judge of their respective strengths from Eq.~(\ref{xilambda}).

A prominent feature of Eqs.~(\ref{MIA1}, \ref{MIA2}, \ref{MIA3}) is the
occurrence of several CP-violating phases, not related to the CKM one. Indeed,
$y_{2}$, $w_{1}$, $\tilde{a}_{4}$, $y_{5}$, $w_{2}$, $w_{4}$ can be complex,
while all the other coefficients have to be real to satisfy the hermiticity of
the squark mass terms. Though it is known that in the MSSM, MFV implies the
presence of new CP-violating phases, up to now they have been considered only
in the trilinear couplings. We see here that the parameter $y_{2}^{\ast}$
brings in an additional CP-violating phase in the LL sector also\footnote{The
impact of having a complex $y_{2}$ was partially analyzed in
Ref.~\cite{EllisLP07}. Indeed, in that work, though universality is imposed on
the squark mass terms, so there is no new CP-phases, each trilinear term has a
CP-phase at the GUT scale. As we will explore in some details later, the
squark mass terms then can develop imaginary parts of precisely the MFV form
by running down to the electroweak scale.}. Though it is competitive only for
sufficiently large $\tan\beta$, since $y_{2}\sim\mathcal{O}(\lambda
^{6}t_{\beta}^{2})$, it could nevertheless play a role in $b\rightarrow s$ and
$b\rightarrow d$ transitions, but not in $s\rightarrow d$ ones (which are in
any case very constrained by $\varepsilon_{K}$). Since $(\delta_{LL})^{23}$
and $(\delta_{LL})^{13}$ are proportional to each other, it is not clear if,
for example, this phase could explain the tension observed recently by the fit
to the $b\rightarrow s$ transitions done in Ref.~\cite{Bona2008}. Before
drawing any conclusion about the validity of the MFV hypothesis at low energy,
the role of these phases in the phenomenology should be investigated in detail.

Concerning the CP-violating phases in the trilinear couplings, they can also
have an impact on low energy observables, though mostly for $b\rightarrow s$
and $b\rightarrow d$ transitions. Indeed, in the $s\rightarrow d$ sector,
given the suppression of $w_{2}$, even at large $\tan\beta$, the quadratic
mass-insertions of the form $(\delta_{RL}^{U})^{32}(\delta_{RL}^{U})^{31\ast}$
dominates \cite{ColangeloI98}. The CP-phase of $\tilde{a}_{4}$ thus plays no
role, and these transitions are entirely tuned by the CKM phase (this remains
true when the contribution of the $\mu$-term to the LR mass insertion is
added, see Ref.~\cite{MFVpheno1} for more details). This is a priori not
trivial looking back at the MFV parametrization (\ref{eq:inter1}), because of
the presence of the complex $b_{i}$ and $c_{i}$ terms in the expansion of
$\mathbf{A}^{U}$. However, the contributions of these terms is always
accompanied with light-quark masses, and is thus suppressed~\cite{MFVpheno1}.
This suppression is immediately visible in our parametrization, based on the
counting rules (\ref{eq:counting}).

\subsection{Lepton sector}

Since the RGE's for the (s)quark parameters depend also on the (s)lepton
sector, we add%
\begin{align}
\mathbf{m}_{L}^{2}  &  =m_{0}^{2}\left[  a_{6}+b_{13}\mathbf{Y}_{e}^{\dagger
}\mathbf{Y}_{e}\right]  \;,\;\nonumber\\
\mathbf{m}_{E}^{2}  &  =m_{0}^{2}\left[  a_{7}+b_{14}\mathbf{Y}_{e}%
\mathbf{Y}_{e}^{\dagger}\right]  \;,\;\nonumber\\
\mathbf{A}^{E}  &  =A_{0}\mathbf{Y}_{e}\left[  a_{8}+b_{15}\mathbf{Y}%
_{e}^{\dagger}\mathbf{Y}_{e}\right]  \;.
\end{align}
For simplicity, we do not allow for flavour mixing in the lepton sector, and
the counting rules describing the hierarchical lepton masses are also in
powers of $\lambda$ as $m_{\tau}/m_{t}\sim\mathcal{O}(\lambda^{3})$, $m_{\mu
}/m_{t}\sim\mathcal{O}(\lambda^{5})$, $m_{e}/m_{t}\sim\mathcal{O}(\lambda
^{7})$. Projecting the soft SUSY breaking matrices in the leptonic sector onto
the $X_{i}$ basis, we get
\begin{equation}
\mathbf{m}_{L}^{2}=\tilde{a}_{6}+y_{6}X_{1}\;,\quad\mathbf{m}_{E}%
^{2}=\tilde{a}_{7}+y_{7}X_{1}\;,\quad\mathbf{A}^{L}=\tilde{a}_{8}X_{1}%
+w_{5}X_{2}\;,
\end{equation}
Relative to the leading terms, $\tilde{a}_{8}$ (which incorporates $y_{\tau}$)
as well as the $y_{i}$'s can become of order one if $\tan\beta\sim\lambda
^{-3}$:
\begin{equation}
\frac{\tilde{a}_{8}}{\tilde{a}_{4}}\sim\mathcal{O}(\lambda^{3}t_{\beta
})\;,\;\;\;\frac{y_{6}}{\tilde{a}_{6}}\sim\frac{y_{7}}{\tilde{a}_{7}}%
\sim\mathcal{O}(\lambda^{6}t_{\beta}^{2})\;,\;\;\;\frac{w_{5}}{\tilde{a}_{8}%
}\sim\mathcal{O}(\lambda^{2})\;. \label{xilambda2}%
\end{equation}
Also in this sector, this representation remains valid in the case of large
$\tan\beta$, up to $\tan\beta\sim\mathcal{O}(m_{t}/m_{b})$) -- on the other
hand, if the latter is not large, $y_{6,7}$ can be dropped.

\section{Derivation of the renormalization group equations}

\label{sec:RGE} We have shown before that imposing the principle of MFV to the
soft SUSY-breaking terms is, from a mathematical point of view, a mere
reparametrization of these. One can therefore do an exact rewriting of the
RGE's for the soft SUSY-breaking matrices into RGE's for the MFV coefficients,
the $z_{i}$'s appearing in Eq.~(\ref{eq:mq2g}). If the RG evolution does not
make any of the $z_{i}$'s become huge, then we can safely drop also from the
RGE's the irrelevant or redundant coefficients and get the RGE's for the
reduced MFV parameters, the $x_{i}$'s of Eq.~(\ref{eq:xMFV}). In the
following, we discuss this procedure in details, and give the RGE's for the
reduced set of MFV parameters. The first question we have to address, however,
is whether to allow our basis matrices (either the $\mathbf{Y}_{u,d}$ and
products thereof, or the $X_{i}$) to run or not. In order to do this, we first
have to discuss how the Yukawa matrices (and correspondingly, the CKM matrix) run.

\subsection{RGE for the Yukawa matrices}

We first analyze the case of moderate $\tan\beta$, and discuss below the
necessary modifications when $\tan\beta$ becomes large. Our starting point are
the counting rules for the quark masses which we have defined in
Eq.~(\ref{eq:counting}), the background values of the Yukawa matrices given in
Eq.~(\ref{eq:Ybackv}) and the Wolfenstein parametrization~(\ref{eq:CKMW}) for
the CKM matrix. We set the up-quark, down-quark and electron masses equal to
zero and systematically neglect anything of order $\lambda^{6}$ or higher. The
Yukawa couplings at the electroweak scale then have the following forms:%
\begin{equation}
\mathbf{Y}_{u}\left(  M_{Z}\right)  =y_{c}X_{6}+y_{t}X_{5}\;,\;\mathbf{Y}%
_{d}\left(  M_{Z}\right)  =y_{s}X_{2}+y_{b}X_{1}\;,\;\mathbf{Y}_{e}\left(
M_{Z}\right)  =y_{\mu}X_{2}+y_{\tau}X_{1}\;\;. \label{eq:Ysa}%
\end{equation}
For very large $\tan\beta$, these initial conditions may have to be amended,
as discussed below.

In the RGE for the Yukawa matrices themselves, but also of other matrices of
the MSSM, products of several Yukawa matrices appear. But since the $X_{i}$
basis is closed under matrix multiplication, all these RGE's corrections can
be projected back on the $X_{i}$ basis. Applied to the Yukawa matrices
themselves, if we run according to the RGE's of the MSSM with the initial
conditions in Eq.~(\ref{eq:Ysa}), we find out (as expected) that additional
structures appear. But once our counting rules in $\lambda$ are enforced on
the RGE also\footnote{In view of the loop factor $1/(16\pi^{2})$ in the
definition of the beta functions, we keep leading terms only up to order
$\lambda^{4}$ in the beta functions, in contrast to $\lambda^{6}$ in the
matrices.}, it turns out that it is sufficient to add only one term for
$\mathbf{Y}_{u}$ and $\mathbf{Y}_{d}$ to obtain an RGE invariant structure:%
\begin{align}
\mathbf{Y}_{u}(\mu)  &  =y_{c}(\mu)X_{6}+y_{t}(\mu)X_{5}+c_{t}(\mu
)X_{1}\;,\nonumber\\
\mathbf{Y}_{d}(\mu)  &  =y_{s}(\mu)X_{2}+y_{b}(\mu)X_{1}+c_{b}(\mu
)X_{5}\;\;,\nonumber\\
\mathbf{Y}_{e}(\mu)  &  =y_{\mu}(\mu)X_{2}+y_{\tau}(\mu)X_{1}\;\;\mathbf{,}
\label{eq:Yusc}%
\end{align}
The running of the three $3\times3$ Yukawa matrices collapses to that of only
$8$ independent parameters.

We stress that the matrices $X_{i}$ are held fixed -- of course, the physical
CKM matrix runs also, as one can easily realize by rediagonalizing
$\mathbf{Y}_{u,d}\left(  \mu\right)  $ in Eq.~(\ref{eq:Yusc}) with
$c_{t},c_{b}\neq0$. The CKM matrix at the electroweak scale is however given
once and for all, and we use it to define a basis of numerical matrices,
Eq.~(\ref{eq:Mbasis}), in which to express the running Yukawa couplings. One
can think of this basis as a fixed grid on which the RGE's for all
flavour-breaking parameters are projected. As will become clear in the
following, this grid is particularly well-suited to enforce the MFV counting
rules on the RGE, as it will permit to separate the rapid flavour-blind
evolutions from the much slower generation of flavour-breaking effects through
the running.

The beta functions of the coefficients in Eq.~(\ref{eq:Yusc}), defined
according to
\begin{equation}
\frac{dC}{dt}=\frac{1}{N}\beta_{C}\;,\qquad N\equiv16\pi^{2}\;,
\end{equation}
with $t=\ln Q/Q_{0}$, then read ($\bar{y}_{t}\equiv y_{t}+c_{t}$, $\bar{y}%
_{b}\equiv y_{b}+c_{b}$)%
\begin{equation}%
\begin{array}
[c]{ll}%
\beta_{y_{t}}=y_{t}\left(  6\bar{y}_{t}^{2}-K_{u}\right)  +c_{b}\bar{y}%
_{b}\bar{y}_{t}\;,\; & \beta_{c_{t}}=c_{t}\left(  6\bar{y}_{t}^{2}%
-K_{u}\right)  +y_{b}\bar{y}_{b}\bar{y}_{t}\;,\;\smallskip\\
\beta_{y_{b}}=y_{b}\left(  6\bar{y}_{b}^{2}+y_{\tau}^{2}-K_{d}\right)
+c_{t}\bar{y}_{t}\bar{y}_{b}\;,\;\; & \beta_{c_{b}}=c_{b}\left(  6\bar{y}%
_{b}^{2}+y_{\tau}^{2}-K_{d}\right)  +\bar{y}_{b}y_{t}\bar{y}_{t}%
\;,\smallskip\\
\beta_{y_{c}}=y_{c}\left(  3\bar{y}_{t}^{2}-K_{u}\right)  \;, & \beta
_{y_{\tau}}=y_{\tau}\left(  4y_{\tau}^{2}+3\bar{y}_{b}^{2}-K_{e}\right)
\;,\smallskip\\
\beta_{y_{s}}=y_{s}\left(  3\bar{y}_{b}^{2}+y_{\tau}^{2}-K_{d}\right)  \;, &
\beta_{y_{\mu}}=y_{\mu}\left(  3\bar{y}_{b}^{2}+y_{\tau}^{2}-K_{e}\right)  \;.
\end{array}
\label{eq:ysRGE}%
\end{equation}
where%
\begin{equation}
K_{u}=\frac{16}{3}g_{3}^{2}+3g_{2}^{2}+\frac{13}{15}g_{1}^{2}\;,\;\;K_{d}%
=K_{u}-\frac{2}{5}g_{1}^{2}\;,\;\;K_{e}=3g_{2}^{2}+\frac{9}{5}g_{1}^{2}\;\;.
\end{equation}
The coefficients $c_{t}$ and $c_{b}$ are zero by definition at $\mu=M_{Z}$,
but are generated by the running at any other scale (see
Fig.~\ref{fig:YukawaRGE} in Appendix~\ref{AppendixRGE}). The leading terms in
their beta-functions are $y_{b}^{2}y_{t}$ and $y_{t}^{2}y_{b}$, respectively,
and from these we can infer (albeit with some degree of arbitrariness) how to
count them. In the following we adopt as counting rule:
\begin{equation}
c_{t}\sim\mathcal{O}(y_{b}^{2}\lambda^{2})\quad\mbox{and}  \quad c_{b}%
\sim\mathcal{O}(y_{b}\lambda^{2})\;\;. \label{cbt}%
\end{equation}
According to this, the particularly simple beta-functions in
Eq.~(\ref{eq:ysRGE}) are accurate up to corrections of order $\lambda^{4}$
(and in many cases to better than this). We checked numerically that indeed,
Eq.~(\ref{eq:ysRGE}) agrees with the full one-loop running of the Yukawa
couplings, after projecting them back on the basis of Eq.~(\ref{eq:Yusc}), to
better than $0.1\%$ for moderate $\tan\beta$.

At large $\tan\beta$, the Yukawa couplings at the electroweak scale are not
always simply related to the quark masses, and the initial conditions at
$\mu=M_{Z}$ in Eq.~(\ref{eq:Ysa}) have to be corrected. Indeed,
non-holomorphic Higgs couplings arise beyond leading order from the combined
breakings of the $U(1)_{PQ}$ symmetry by the $\mu$ and $b$ terms, and of
supersymmetry itself by the soft-breaking terms \cite{HallRS94}. After
electroweak symmetry breaking, non-flavour diagonal, $\tan\beta$-enhanced
contributions to the down-quark mass matrix emerge. The net effect is that in
the basis in which $\mathbf{Y}_{u}$ is given in terms of the physical up-quark
masses and CKM matrix, $\mathbf{Y}_{d}$ is not diagonal:%
\begin{equation}
\mathbf{Y}_{u}=\mathbf{\lambda}_{u}V\;,\;\;\mathbf{Y}_{d}=\mathbf{\lambda}%
_{d}\left(  \mathbf{1}-\mathbf{\Delta}\right)  ^{-1}\;,\;
\label{LTB:background}%
\end{equation}
where $v_{u}\mathbf{\lambda}_{u}$, $v_{d}\mathbf{\lambda}_{d}$ are the
(diagonal) physical quark mass matrices, and $V$ the physical CKM
matrix\footnote{This quark basis is different from the one of e.g.
Ref.~\cite{D'Ambrosio:2002ex}, where the background value for $\mathbf{Y}_{d}$
is kept diagonal, but at the cost of having $\mathbf{Y}_{u}=\mathbf{\lambda
}_{u}V^{\prime}$ with $V^{\prime}$ different from the CKM matrix. Here, the
down quark fields are already mass-eigenstates, since once loop corrections
are added, $\mathbf{M}_{d}=v_{d}\mathbf{Y}_{d}\left(  \mathbf{1}%
-\mathbf{\Delta}\right)  =v_{d}\mathbf{\lambda}_{d}$.}. All the loop-induced,
$\tan\beta$-enhanced corrections are in $\mathbf{\Delta}$, which has the MFV
expansion (under the assumption that $\mathbf{Y}_{d}$ is still sufficiently
hierarchical)%
\begin{equation}
\mathbf{\Delta}=\tan\beta\left(  \varepsilon_{0}\mathbf{1}+\varepsilon
_{1}\mathbf{Y}_{u}^{\dagger}\mathbf{Y}_{u}+\varepsilon_{2}\mathbf{Y}%
_{d}^{\dagger}\mathbf{Y}_{d}+\varepsilon_{3}\mathbf{Y}_{d}^{\dagger}%
\mathbf{Y}_{d}\mathbf{Y}_{u}^{\dagger}\mathbf{Y}_{u}+\varepsilon_{4}%
\mathbf{Y}_{u}^{\dagger}\mathbf{Y}_{u}\mathbf{Y}_{d}^{\dagger}\mathbf{Y}%
_{d}\right)  \;. \label{LTB:delta}%
\end{equation}
The $\varepsilon_{i}$ parameters are loop suppressed and depend on the
mass-spectrum of the specific MSSM model under consideration. However, for
large $\tan\beta$, this suppression can be compensated, at least partially,
leading to a relation between $\mathbf{Y}_{d}$ and $\mathbf{\lambda}_{d}$ with
corrections which may become of $\mathcal{O}(1)$ if $\varepsilon_{i}\tan
\beta\sim1$.

To see what could happen in that case, we remark first that the
Cayley-Hamilton identity (\ref{eq:CH1}) implies
\begin{equation}
\mathbf{X}^{-1}=\frac{6\mathbf{X}^{2}-6\left\langle \mathbf{X}\right\rangle
\mathbf{X}+3\left(  \left\langle \mathbf{X}\right\rangle ^{2}-\left\langle
\mathbf{X}^{2}\right\rangle \right)  }{2\left\langle \mathbf{X}^{3}%
\right\rangle -3\left\langle \mathbf{X}\right\rangle \left\langle
\mathbf{X}^{2}\right\rangle +\left\langle \mathbf{X}\right\rangle ^{3}}\;.
\end{equation}
For $\mathbf{X}=\mathbf{1}-\mathbf{\Delta}$, this becomes%
\begin{equation}
\left(  \mathbf{1}-\mathbf{\Delta}\right)  ^{-1}=\eta_{1}\mathbf{1}+\eta
_{2}\mathbf{\Delta}+\eta_{3}\mathbf{\Delta}^{2}\mathbf{\;}, \label{LTB:series}%
\end{equation}
with the coefficients%
\begin{align}
\eta_{1}/\eta_{3}  &  =1-\left\langle \mathbf{\Delta}\right\rangle
+\frac{1}{2}\left(  \left\langle \mathbf{\Delta}\right\rangle ^{2}%
-\left\langle \mathbf{\Delta}^{2}\right\rangle \right)  ,\;\;\;\;\eta_{2}%
/\eta_{3}=1-\left\langle \mathbf{\Delta}\right\rangle \;,\\
\eta_{3}^{-1}  &  =1-\left\langle \mathbf{\Delta}\right\rangle +\frac{1}%
{2}\left(  \left\langle \mathbf{\Delta}\right\rangle ^{2}-\left\langle
\mathbf{\Delta}^{2}\right\rangle \right)  +\frac{1}{2}\left\langle
\mathbf{\Delta}^{2}\right\rangle \left\langle \mathbf{\Delta}\right\rangle
-\frac{1}{3}\left\langle \mathbf{\Delta}^{3}\right\rangle -\frac{1}%
{6}\left\langle \mathbf{\Delta}\right\rangle ^{3}\;.\nonumber
\end{align}
There is no approximation in this formula: Cayley-Hamilton allows to
completely resum the geometric series expansion of $\left(  \mathbf{1}%
-\mathbf{\Delta}\right)  ^{-1}$. When the mass spectrum is such that
$\varepsilon_{i}\tan\beta$ is small, these formula can be expanded to first
order, leading to $\mathbf{Y}_{d}\approx\mathbf{\lambda}_{d}\left(
\mathbf{1}+\mathbf{\Delta}\right)  $.

Provided the $\varepsilon_{i}$ are not too large, the coefficients $\eta_{i}$
are $\mathcal{O}(1)$ numbers. In that case, the expansion (\ref{LTB:delta}) is
certainly valid, and all the $\tan\beta$-enhanced corrections can be absorbed
as shifts in the values of $y_{b}\left(  M_{Z}\right)  $, $y_{s}\left(
M_{Z}\right)  $ and $c_{b}\left(  M_{Z}\right)  $, since $\left(
\mathbf{1}-\mathbf{\Delta}\right)  ^{-1}$ has the same form as the RGE
effects. We stress that, except for very large $\varepsilon_{i}$, the counting
rules are not upset by these shifts. Only $c_{b}$ tends to become somewhat
larger than before (see Appendix~\ref{AppendixRGE} for a numerical
application), but this increase is quite mild, and the RGE's,
Eq.~(\ref{eq:ysRGE}), remain accurate to better than $1\%$.

In the present paper, our aim is to probe the behavior of the RGE's also in
the large $\tan\beta$ scenario. Since the $\varepsilon_{i}$ depend on the
mass-spectrum, which itself depends on the matching conditions at the
electroweak scale, and hence on the $\varepsilon_{i}$, this problem can be
properly solved only by setting up an iterative procedure, and this is beyond
our scope. Our analysis could in fact be viewed as the first of these
iterations, allowing to derive an approximate mass spectrum, and to compute
the $\varepsilon_{i}$ parameters. Even if in the end the $\varepsilon_{i}$
will turn out to be very large, such that $\eta_{i}\gg1$, we stress that the
counting rule method developed here and in the previous section could be
generalized to more complicated initial conditions, simply by allowing for
additional structures (and parameters) in the Yukawa couplings in
Eq.~(\ref{eq:Ysa}).

\subsection{RGE for the MFV parameters}

We now consider the soft SUSY breaking terms. The beta functions for them are
known (even up to two loops, \cite{Martin:1993zk}) and can also be projected
onto the $X_{i}$ basis. In principle, one could generate new structures beyond
those given in Eq.~(\ref{eq:xMFV}) if the beta functions of some of them were
of a different order than the coefficients themselves. We have verified that
this does not happen, and that the structure in Eq.~(\ref{eq:xMFV}) is indeed
RGE invariant if we stick to our counting rules (\emph{i.e.} deviations from
this structure arise in the running from contributions to the beta functions
which are suppressed by at least $\lambda^{4}$). We stress that this is only a
necessary (but not sufficient) condition to ensure that the MFV hypothesis is
RGE invariant -- an exponential growth can of course be generated by a beta
function which is of the same order as the parameter itself. We will then have
to study the behaviour of the solutions of the RGE in order to reach the
desired conclusion.

The beta functions of the new coefficients turn out to be remarkably simple.
They read for the squark soft-breaking terms:%
\begin{equation}%
\begin{array}
[c]{ll}%
\mathbf{m}_{Q}^{2}: & \beta_{\tilde{a}1}=-\frac{32}{3}g_{3}^{2}|M_{3}%
|^{2}-6g_{2}^{2}|M_{2}|^{2}-\frac{2}{15}g_{1}^{2}|M_{1}|^{2}+\frac{1}{5}%
g_{1}^{2}S\;,\smallskip\\
& \beta_{x_{1}}=2y_{t}^{2}\left(  m_{Hu}^{2}+\tilde{a}_{1}+\tilde{a}_{2}%
+x_{1}+x_{2}+\operatorname{Re}y_{2}\right)  +2\left(  |\tilde{a}_{4}%
|^{2}+|y_{5}|^{2}\right)  \;,\smallskip\\
& \beta_{y_{1}}=2y_{b}^{2}\left(  m_{Hd}^{2}+\tilde{a}_{1}+\tilde{a}_{3}%
+y_{1}+\operatorname{Re}y_{2}+y_{3}\right)  +2\left(  |\tilde{a}_{5}%
|^{2}+|y_{4}|^{2}\right)  \;,\smallskip\\
& \beta_{y_{2}}=y_{b}^{2}(x_{1}+y_{2})+y_{t}^{2}\left(  y_{1}+y_{2}\right)
+2\left(  \tilde{a}_{5}^{\ast}y_{5}+\tilde{a}_{4}y_{4}^{\ast}\right)
\;,\medskip\\
\mathbf{m}_{U}^{2}: & \beta_{\tilde{a}2}=-\frac{32}{3}g_{3}^{2}|M_{3}%
|^{2}-\frac{32}{15}g_{1}^{2}|M_{1}|^{2}-\frac{4}{5}g_{1}^{2}S\;,\smallskip\\
& \beta_{x_{2}}=4y_{t}^{2}\left(  m_{Hu}^{2}+\tilde{a}_{1}+\tilde{a}_{2}%
+x_{1}+x_{2}+y_{1}+2\operatorname{Re}y_{2}\right)  +4|\tilde{a}_{4}+y_{4}%
|^{2}\;,\smallskip\\
\mathbf{m}_{D}^{2}: & \beta_{\tilde{a}3}=-\frac{32}{3}g_{3}^{2}|M_{3}%
|^{2}-\frac{8}{15}g_{1}^{2}|M_{1}|^{2}+\frac{2}{5}g_{1}^{2}S\;,\smallskip\\
& \beta_{y_{3}}=4y_{b}^{2}\left(  m_{Hd}^{2}+\tilde{a}_{1}+\tilde{a}_{3}%
+x_{1}+y_{1}+2\operatorname{Re}y_{2}+y_{3}\right)  +4|\tilde{a}_{5}+y_{5}%
|^{2}\;,\smallskip\\
& \beta_{w_{1}}=2w_{1}\bar{y}_{b}^{2}+4\left(  \tilde{a}_{5}+y_{5}\right)
w_{4}^{\ast}-4A\lambda^{2}\left(  \left(  x_{1}+y_{2}\right)  \bar{y}_{b}%
y_{s}+w_{3}^{\ast}y_{5}\right)  \;,\medskip\\
\mathbf{A}^{U}: & \beta_{\tilde{a}_{4}}=\tilde{a}_{4}\left(  18y_{t}^{2}%
-K_{u}\right)  +y_{t}\left(  11y_{4}y_{t}+2y_{5}y_{b}+K_{u}^{\prime}\right)
\;,\smallskip\\
& \beta_{y_{4}}=y_{4}\left(  y_{b}^{2}+7y_{t}^{2}-K_{u}\right)  +\tilde{a}_{4}%
y_{b}^{2}+2\tilde{a}_{5}y_{b}y_{t}\;,\smallskip\\
& \beta_{w_{2}}=w_{2}\left(  3y_{t}^{2}-K_{u}\right)  +y_{c}\left(
6y_{t}\left(  \tilde{a}_{4}+y_{4}\right)  +K_{u}^{\prime}\right)
\;,\medskip\\
\mathbf{A}^{D}: & \beta_{\tilde{a}_{5}}=\tilde{a}_{5}\left(  18y_{b}%
^{2}+y_{\tau}^{2}-K_{d}\right)  +y_{b}\left(  11y_{5}y_{b}+2y_{4}%
y_{t}+2\tilde{a}_{8}y_{\tau}+K_{d}^{\prime}\right)  \;,\smallskip\\
& \beta_{y_{5}}=y_{5}\left(  7y_{b}^{2}+y_{t}^{2}+y_{\tau}^{2}-K_{d}\right)
+\tilde{a}_{5}y_{t}^{2}+2\tilde{a}_{4}y_{b}y_{t}+c_{b}K_{d}^{\prime
}\;,\smallskip\\
& \beta_{w_{3}}=w_{3}\left(  3y_{b}^{2}+y_{\tau}^{2}-K_{d}\right)
+y_{s}\left(  6y_{b}\left(  \tilde{a}_{5}+y_{5}\right)  +2\tilde{a}_{8}%
y_{\tau}+K_{d}^{\prime}\right)  \;,\smallskip\\
& \beta_{w_{4}}=w_{4}\left(  8y_{b}^{2}+y_{t}^{2}+y_{\tau}^{2}-K_{d}\right)
-A\lambda^{2}\left(  2y_{s}y_{t}\left(  \tilde{a}_{4}+y_{4}\right)
+w_{3}y_{t}^{2}\right)  \;,
\end{array}
\label{RGE1}%
\end{equation}
and for the slepton soft-breaking terms:%
\begin{equation}%
\begin{array}
[c]{ll}%
\mathbf{m}_{L}^{2}: & \beta_{\tilde{a}6}=-6g_{2}^{2}|M_{2}|^{2}-\frac{6}%
{5}g_{1}^{2}|M_{1}|^{2}-\frac{3}{5}g_{1}^{2}S\;,\smallskip\\
& \beta_{y_{6}}=2y_{\tau}^{2}\left(  m_{Hd}^{2}+\tilde{a}_{6}+\tilde{a}_{7}%
+y_{6}+y_{7}\right)  +2|\tilde{a}_{8}|^{2}\;,\medskip\\
\mathbf{m}_{E}^{2}: & \beta_{\tilde{a}7}=-\frac{24}{5}g_{1}^{2}|M_{1}%
|^{2}+\frac{6}{5}g_{1}^{2}S\;,\smallskip\\
& \beta_{y_{7}}=4y_{\tau}^{2}\left(  m_{Hd}^{2}+\tilde{a}_{6}+\tilde{a}_{7}%
+y_{6}+y_{7}\right)  +4|\tilde{a}_{8}|^{2}\;,\medskip\\
\mathbf{A}^{E}: & \beta_{\tilde{a}_{8}}=\tilde{a}_{8}\left(  12y_{\tau}%
^{2}+3\bar{y}_{b}^{2}-K_{e}\right)  +y_{\tau}\left(  6\left(  \tilde{a}_{5}%
+x_{10}\right)  \bar{y}_{b}+K_{e}^{\prime}\right)  \;,\smallskip\\
& \beta_{w_{5}}=w_{5}\left(  y_{\tau}^{2}+3\bar{y}_{b}^{2}-K_{e}\right)
+y_{\mu}\left(  6\left(  \tilde{a}_{5}+y_{5}\right)  \bar{y}_{b}%
+2\tilde{a}_{8}y_{\tau}+K_{e}^{\prime}\right)  \;,
\end{array}
\label{RGE2}%
\end{equation}
where $V_{cb}\simeq A\lambda^{2}\sim4\cdot10^{-2}$ is the element of the CKM
matrix, and where we have introduced the abbreviations%
\begin{gather}
K_{u}^{\prime}=\frac{32}{3}g_{3}^{2}M_{3}+6g_{2}^{2}M_{2}+\frac{26}{15}%
g_{1}^{2}M_{1},\;\;K_{d}^{\prime}=K_{u}^{\prime}+\frac{4}{5}g_{1}^{2}%
M_{1},\;K_{e}^{\prime}=6g_{2}^{2}M_{2}+\frac{18}{5}g_{1}^{2}M_{1}\;,\\
S=m_{H_{u}}^{2}-m_{H_{d}}^{2}+3\left(  \tilde{a}_{1}-2\tilde{a}_{2}%
+\tilde{a}_{3}-\tilde{a}_{6}+\tilde{a}_{7}\right)  +x_{1}-2x_{2}%
+y_{1}+2\operatorname{Re}y_{2}+y_{3}-y_{6}+y_{7}\;.\nonumber
\end{gather}
The only other RGE's depending on the sfermion soft-breaking terms are the
Higgs parameters, which now take the form%
\begin{align}
\beta_{\mu}  &  =\mu\left(  3\bar{y}_{t}^{2}+3\bar{y}_{b}^{2}+y_{\tau}%
^{2}+3y_{s}^{2}+y_{\mu}^{2}-3g_{2}^{2}-\tfrac{3}{5}g_{1}^{2}\right)
\;,\nonumber\\
\beta_{b}  &  =b\left(  3\bar{y}_{t}^{2}+3\bar{y}_{b}^{2}+y_{\tau}^{2}%
+3y_{s}^{2}+y_{\mu}^{2}-3g_{2}^{2}-\tfrac{3}{5}g_{1}^{2}\right) \nonumber\\
&  \;\;\;\;+\mu\left(  6\left(  \tilde{a}_{4}+y_{4}\right)  \bar{y}%
_{t}+6\left(  \tilde{a}_{5}+y_{5}\right)  \bar{y}_{b}+2\tilde{a}_{8}y_{\tau
}+6w_{3}y_{s}+2w_{5}y_{\mu}+6g_{2}^{2}M_{2}+\tfrac{6}{5}g_{1}^{2}M_{1}\right)
\;,\nonumber\\
\beta_{m_{H_{u}}^{2}}  &  =6\bar{y}_{t}^{2}\left(  m_{H_{u}}^{2}%
+\tilde{a}_{1}+\tilde{a}_{2}+x_{1}+x_{2}+y_{1}+2\operatorname{Re}y_{2}\right)
+6|\tilde{a}_{4}+y_{4}|^{2}\nonumber\\
&  \;\;\;\;-6g_{2}^{2}|M_{2}|^{2}-\frac{6}{5}g_{1}^{2}|M_{1}|^{2}+\frac{3}%
{5}g_{1}^{2}S+6y_{c}^{2}\left(  m_{H_{u}}^{2}+\tilde{a}_{1}+\tilde{a}_{2}%
\right)  +6|w_{2}|^{2}\;,\nonumber\\
\beta_{m_{H_{d}}^{2}}  &  =6\bar{y}_{b}^{2}\left(  m_{H_{d}}^{2}%
+\tilde{a}_{1}+\tilde{a}_{3}+x_{1}+y_{1}+2\operatorname{Re}y_{2}+y_{3}\right)
+6|\tilde{a}_{5}+y_{5}|^{2}\nonumber\\
&  \;\;\;\;-6g_{2}^{2}|M_{2}|^{2}-\frac{6}{5}g_{1}^{2}|M_{1}|^{2}-\frac{3}%
{5}g_{1}^{2}S+2y_{\tau}^{2}\left(  m_{H_{d}}^{2}+\tilde{a}_{6}+\tilde{a}_{7}%
+y_{6}+y_{7}\right)  +2|\tilde{a}_{8}|^{2}\nonumber\\
&  \;\;\;\;+6y_{s}^{2}\left(  m_{H_{d}}^{2}+\tilde{a}_{1}+\tilde{a}_{3}%
\right)  +6|w_{3}|^{2}+2y_{\mu}^{2}\left(  m_{H_{d}}^{2}+\tilde{a}_{6}%
+\tilde{a}_{7}\right)  +2|w_{5}|^{2}\;. \label{RGE3}%
\end{align}
In these last equations, corrections of relative $\mathcal{O}(\lambda^{4})$
were kept. They can be relevant at the percent level in some corner of
parameter space because of their rather fast running.

We stress that the RGE's (\ref{RGE1}, \ref{RGE2}, \ref{RGE3}) are to be taken
as they are in the case of large $\tan\beta$, \emph{i.e.} when $t_{\beta}%
\sim\mathcal{O}(\lambda^{-3})$, as long as the Yukawa couplings have the
structure shown in Eq.~(\ref{eq:Yusc}). On the other hand, for $\tan\beta$ of
order $\lambda^{-2}$ or even $\lambda^{-1}$, some couplings and the
corresponding beta functions change their order and may become negligibly
small. In that case, the flavour structure of the theory as well as the RGE's
become significantly simpler.

\section{Running MFV in the moderate $\tan\beta$ case}

In this and the following section, we will illustrate the behaviour of the
various parameters as functions of the scale with the help of a few examples.
We first discuss the case of moderate $\tan\beta$, and then the case of a
large one. In order to do this, we take as reference two of the Snowmass
benchmark points \cite{Allanach:2002nj}, and use the MFV parameters to explore
flavour violations around these points. These numerical examples are essential
in order to understand the solutions of the RGE's. In case of moderate
$\tan\beta$, however, the structure of the MFV MSSM, and of the corresponding
RGE's, simplifies so much that one can provide a semi-analytical solution of
the RGE's. We will now first derive these, then present the numerics, and show
how one can understand the observed behaviour with the help of the analytical formulae.

\subsection{Analytical solutions}

\subsubsection{Solutions for the $\tilde{a}_{i}$'s}

We first observe that the term proportional to $S$ in the RGE's for
$\tilde{a}_{1,2,3}$ is typically very small: it is multiplied by the small
gauge coupling $g_{1}^{2}$ and a small coefficient, and the combination of
massive coupling constants which appears in there is zero at the GUT scale for
initial conditions of the mSUGRA type. The numerical examples we will discuss
below will show that this term can indeed be safely neglected, unless one
takes very special initial conditions. Dropping $S$, the RGE's for
$\tilde{a}_{1,2,3}$ admit a simple explicit analytical solution ($t_{0}\equiv
t(M_{\text{GUT}})$):
\begin{align}
\tilde{a}_{1}(t)  &  =\tilde{a}_{1}(t_{0})+\frac{8}{9}\left(  |M_{3}%
(t)|^{2}-|M_{3}(t_{0})|^{2}\right)  -\frac{3}{2}\left(  |M_{2}(t)|^{2}%
-|M_{2}(t_{0})|^{2}\right) \nonumber\\
&  \;\;\;\;-\frac{1}{198}\left(  |M_{1}(t)|^{2}-|M_{1}(t_{0})|^{2}\right)
\;,\nonumber\\
\tilde{a}_{2}(t)  &  =\tilde{a}_{2}(t_{0})+\frac{8}{9}\left(  |M_{3}%
(t)|^{2}-|M_{3}(t_{0})|^{2}\right)  -\frac{8}{99}\left(  |M_{1}(t)|^{2}%
-|M_{1}(t_{0})|^{2}\right)  \;,\nonumber\\
\tilde{a}_{3}(t)  &  =\tilde{a}_{3}(t_{0})+\frac{8}{9}\left(  |M_{3}%
(t)|^{2}-|M_{3}(t_{0})|^{2}\right)  -\frac{2}{99}\left(  |M_{1}(t)|^{2}%
-|M_{1}(t_{0})|^{2}\right)  \;. \label{eq:a123}%
\end{align}
Typically, the term $|M_{3}(t)|^{2}$, which grows fast when evolving down to
the electroweak scale, will dominate over the rest, so that all three
$\tilde{a}_{i}$'s turn out to be of the order of the gluino mass squared at
the electroweak scale.

The RGE's for $\tilde{a}_{4,5}$ simplify as follows in the moderate $\tan
\beta$ case\footnote{Note that the imaginary parts of $\tilde{a}_{4,5}$ miss
the term proportional to $K_{u,d}^{\prime}$ in the beta function.}:
\begin{align}
\beta_{\tilde{a}_{4}}  &  =\tilde{a}_{4}\left(  18y_{t}^{2}-K_{u}\right)
+y_{t}K_{u}^{\prime}\;,\nonumber\\
\beta_{\tilde{a}_{5}}  &  =\tilde{a}_{5}\left(  18y_{b}^{2}+y_{\tau}^{2}%
-K_{d}\right)  +y_{b}K_{d}^{\prime}\;\;. \label{eq:a4a5}%
\end{align}
Both RGE's are of the form
\begin{equation}
\frac{dA}{dt}=\frac{1}{N}\Big[  f_{A}(t)+A(t)g_{A}(t)\Big]  \;,
\label{eq:mastereq}%
\end{equation}
with $f_{A}(t)$ and $g_{A}(t)$ known functions. The solution of this equation
reads
\begin{equation}
A(t)=\left[  A(t_{0})-\frac{1}{N}\int_{t}^{t_{0}}dt^{\prime}\frac{f_{A}%
(t^{\prime})}{G_{A}(t^{\prime})}\right]  G_{A}(t)\;,\quad\quad G_{A}%
(t)=\exp\left[  -\frac{1}{N}\int_{t}^{t_{0}}dt^{\prime}g_{A}(t^{\prime
})\right]  \;. \label{eq:mastersol}%
\end{equation}

\subsubsection{Solutions for $m_{H_{u,d}}^{2}$ and $x_{1,2}$}

The RGE's for $m_{H_{u}}^{2}$, $x_{1}$ and $x_{2}$ are coupled, even in the
moderate $\tan\beta$ case. In that case, however, one can diagonalize the beta
functions and bring them to the form (\ref{eq:mastereq}), by taking the
following combinations:
\begin{equation}
n_{1}=\frac{1}{2}\left[  m_{H_{u}}^{2}+x_{1}+x_{2}\right]  \;,\quad
n_{2}=\frac{1}{2}\left[  m_{H_{u}}^{2}-x_{1}-x_{2}\right]  \;,\quad
n_{3}=2x_{1}-x_{2}\;.
\end{equation}
The $f$ and $g$ functions for these combinations read
\begin{align}
g_{n_{1}}  &  =12y_{t}^{2}\qquad f_{n_{1}}=-3g_{2}^{2}|M_{2}|^{2}-\frac{3}%
{5}g_{1}^{2}|M_{1}|^{2}+6\left[  y_{t}^{2}(\tilde{a}_{1}+\tilde{a}_{2}%
)+|\tilde{a}_{4}|^{2}\right] \nonumber\\
g_{n_{2}}  &  =0\qquad\;\;\;\;\;f_{n_{2}}=-3g_{2}^{2}|M_{2}|^{2}-\frac{3}%
{5}g_{1}^{2}|M_{1}|^{2}\nonumber\\
g_{n_{3}}  &  =0\qquad\;\;\;\;\;f_{n_{3}}=0\;\;,
\end{align}
and are all known functions. One can plug them into the general solution
(\ref{eq:mastersol}) and gets:
\begin{align}
n_{1}(t)  &  =\left[  n_{1}(t_{0})-\frac{1}{N}\int_{t}^{t_{0}}dt^{\prime
}\frac{f_{n_{1}}(t^{\prime})}{G_{n_{1}}(t^{\prime})}\right]  G_{n_{1}%
}(t)\;,\nonumber\\
n_{2}(t)  &  =n_{2}(t_{0})-\frac{3}{4}\left(  |M_{2}(t)|^{2}-|M_{2}%
(t_{0})|^{2}\right)  -\frac{1}{44}\left(  |M_{1}(t)|^{2}-|M_{1}(t_{0}%
)|^{2}\right)  \;,\nonumber\\
n_{3}(t)  &  =n_{3}(t_{0})\;, \label{eq:n123sol}%
\end{align}
and from these solutions, reconstructs the physical parameters:
\begin{equation}
m_{H_{u}}^{2}=n_{1}+n_{2}\qquad x_{1}=\frac{1}{3}\left[  n_{1}-n_{2}%
+n_{3}\right]  \qquad x_{2}=\frac{1}{3}\left[  2(n_{1}-n_{2})-n_{3}\right]
\;\;. \label{eq:x12sol}%
\end{equation}
The most important feature of these analytical solutions is that the integral
containing the function $f_{n_{1}}$ (which gives a negative contribution)
dominates, because it contains the fast growing functions $\tilde{a}_{i}$,
which behave like the gluino masses. This growth is only partially compensated
by the decreasing function $G_{n_{1}}$ (as $t$ decreases from $t_{0}$ down to
the electroweak scale). The function $n_{2}$, which is positive, grows much
more slowly, approximately linearly in $t_{0}-t$. From these analytical
solutions, one can also clearly see how much the values of these parameters at
any scale depend on their initial conditions at the GUT scale:
\begin{align}
m_{H_{u}}^{2}(t)  &  =\frac{m_{H_{u}}^{2}(t_{0})}{2}\left[  G_{n_{1}%
}(t)+1\right]  +\frac{1}{2}\left(  x_{1}(t_{0})+x_{2}(t_{0})\right)  \left[
G_{n_{1}}(t)-1\right]  +\ldots\nonumber\\
x_{1}(t)  &  =\frac{m_{H_{u}}^{2}(t_{0})}{6}\left[  G_{n_{1}}(t)-1\right]
+\frac{x_{1}(t_{0})}{6}\left[  G_{n_{1}}(t)+5\right]  +\frac{x_{2}(t_{0})}%
{6}\left[  G_{n_{1}}(t)-1\right]  +\ldots\nonumber\\
x_{2}(t)  &  =\frac{m_{H_{u}}^{2}(t_{0})}{3}\left[  G_{n_{1}}(t)-1\right]
+\frac{x_{1}(t_{0})}{3}\left[  G_{n_{1}}(t)-1\right]  +\frac{x_{2}(t_{0})}%
{3}\left[  G_{n_{1}}(t)+2\right]  +\ldots\label{eq:x12ic}%
\end{align}
where terms which are independent from the initial conditions have been
omitted. In order to understand this dependence, one only needs to know the
behaviour of the function $G_{n_{1}}$. As just mentioned, this decreases from
the GUT to the electroweak scale, and in fact does so almost linearly from $1$
to about $0.23$.

For moderate $\tan\beta$, the RGE for $m_{H_{d}}^{2}$ can be solved in terms
of known functions, whose evolutions are not influenced by $m_{H_{d}}^{2}$
itself. The $f$ and $g$ functions read:
\begin{align}
g_{m_{H_{d}}^{2}}  &  =6y_{b}^{2}+2y_{\tau}^{2}\;,\\
f_{m_{H_{d}}^{2}}  &  =6y_{b}^{2}(\tilde{a}_{1}+\tilde{a}_{3}+x_{1})+6y_{\tau
}^{2}(\tilde{a}_{6}+\tilde{a}_{7})+6|\tilde{a}_{5}+y_{5}|^{2}+2|\tilde{a}_{8}%
|^{2}-6g_{2}^{2}|M_{2}|^{2}-\frac{6}{5}g_{1}^{2}|M_{1}|^{2}\;\;.\nonumber
\end{align}

\subsubsection{Solutions for the $y_{i}$'s and the $w_{i}$'s}

The same applies to all the remaining parameters. One can give their solution
in the form of Eq.~(\ref{eq:mastersol}) simply by identifying the $f$ and $g$
functions for each of them. We give here a couple of examples:%
\begin{equation}%
\begin{array}
[c]{ll}%
g_{y_{1}}=2y_{b}^{2}\;, & f_{y_{1}}=2y_{b}^{2}(m_{H_{d}}^{2}+\tilde{a}_{1}%
+\tilde{a}_{3})+2|\tilde{a}_{5}|^{2}\;,\\
g_{y_{2}}=y_{t}^{2}\;, & f_{y_{2}}=y_{b}^{2}x_{1}+y_{t}^{2}y_{1}%
+2(\tilde{a}_{4}y_{4}^{\ast}+\tilde{a}_{5}^{\ast}y_{5})\;.
\end{array}
\end{equation}
In all these cases, the analytical solutions allow one to immediately
understand what depends on what, and in particular, which initial conditions
influence the behaviour of the various parameters at the electroweak scale, as
we are now going to show in the following subsection.

\subsection{Numerical example: the SPS-1a benchmark point}

\label{sec:SPS1a} 
\begin{figure}[ptb]
\includegraphics[width=16cm]{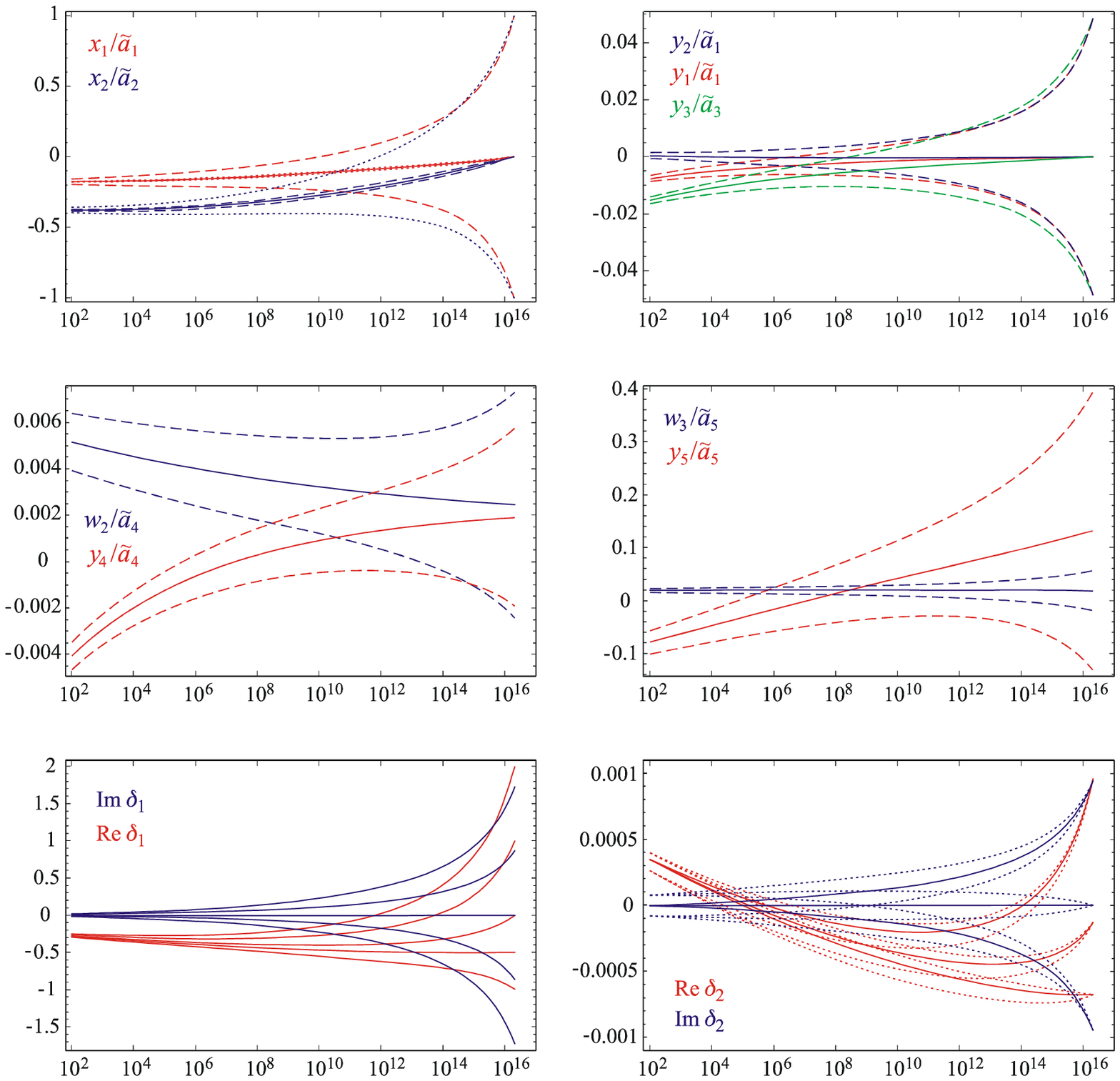}  \caption{Upper four plots: RGE
evolution of the MFV parameters for the SPS-1a benchmark point. The solid
curves always show the evolution with mSUGRA type of initial conditions. In
the upper-left panel, the dashed (dotted) lines show the evolution when
$x_{1(2)}(M_{\text{GUT}})=\pm M_{0}^{2}$. In all other cases, for each
parameter, only three curves are shown -- the upper and lower ones (always
shown as dashed) correspond to different initial conditions for that single
parameter. Lower two plots: the real and imaginary parts of the mass
insertions $\delta_{1(2)}\equiv(\delta_{RL}^{U(D)})^{32}/V_{ts}=(\delta
_{RL}^{U(D)})^{31}/V_{td}$. For these plots, the initial conditions are varied
independently for $\mathbf{A}^{U}$ and $\mathbf{A}^{D}$, allowing for a
CP-phase as $A_{0}=re^{i\phi}$ with $r=0\rightarrow200$ GeV and $\phi$ between
$\pm180{{}^{\circ}}$. Dashed lines in the $\delta_{2}$ plot show the impact of
the initial conditions for $\mathbf{A}^{U}$, while the sensitivity of
$\delta_{1}$ to those for $\mathbf{A}^{D}$ is negligible. The behaviours of
the other mass-insertions, Eqs.~(\ref{MIA1}, \ref{MIA2}, \ref{MIA3}), can
easily be obtained from those of the parameters shown in the upper four plots.
In particular, note that $y_{2}\left(  M_{Z}\right)  \ll x_{1}\left(
M_{Z}\right)  $, both because of the $\tan\beta$ suppression, and because of
RGE effects.}%
\label{fig:sps1}%
\end{figure}

The SPS-1a point is specified by $m_{0}=-A_{0}=100$ GeV and
$m_{1/2}=250$ GeV at the GUT scale, and $\tan\beta=10$. The running of the MFV
parameters in the neighbourhood of the SPS-1a benchmark point is illustrated
in the Figs.~\ref{fig:sps1}. We have checked that the running evaluated by
solving our simplified RGE's gives almost identical results to the running
evaluated with the full one-loop RGE's of Ref.~\cite{Martin:1993zk} (for the
numerical analysis and the figures, we have included also the $\lambda^{2}$
corrections to the beta functions given in Appendix~\ref{app:la2}). We have
solved both sets of differential equations with the same input at the GUT
scale and then compared the full mass matrices at the electroweak scale -- the
differences in all the entries are always consistent with our counting rules,
i.e. of the order of the terms neglected in our simplified RGE's.

The most prominent feature emerging from the figures is that all the
parameters ratios tend to converge to one point at the low scale,
independently of the starting point at the high scale -- \emph{i.e.} they show
a fixed-point kind of behaviour, as recently observed in
Ref.~\cite{Paradisi:2008qh}. We stress, however, that by separating the
leading parameters (the $\tilde{a}_{i}$'s and the $x_{i}$'s) from the
suppressed ones, we can better see the behaviour of the small flavour
violations, governed by the $y_{i}$'s and the $w_{i}$'s. Our analysis shows
that these also converge to fixed points.

\begin{table}[t]%
\begin{tabular}
[c]{|l|l|l|}\hline
& $M_{\text{GUT}}$ & $M_{Z}$\\\hline
$\tilde{a}_{1}/M_{3}^{2}$ & $0.16(1+\Delta_{1})$ & \hskip 0.31 cm
$0.86+0.019\Delta_{1}$\\
$\tilde{a}_{2}/M_{3}^{2}$ & $0.16(1+\Delta_{2})$ & \hskip 0.31 cm
$0.81+0.015\Delta_{2}$\\
$\tilde{a}_{3}/M_{3}^{2}$ & $0.16(1+\Delta_{3})$ & \hskip 0.31 cm
$0.80+0.019\Delta_{3}$\\
$\tilde{a}_{4}/(-y_{t}M_{3})$ & $0.40(1+\Delta_{4})$ & \hskip 0.31 cm
$0.81+0.03\Delta_{4}$\\
$\tilde{a}_{5}/(-y_{b}M_{3})$ & $0.40(1+\Delta_{5})$ & \hskip 0.31 cm
$1.48+0.14\Delta_{5}$\\\hline
$x_{1}/\tilde{a}_{1}$ & $\delta_{1}$ &
$-0.18-0.003\Delta_2-0.01\Delta_{4}+0.02\delta_{1}-0.003\delta_{2}$\\
$y_{1}/\tilde{a}_{1}$ & $\epsilon_{1}t_{\beta}^{2}\lambda^{6}$ &
$(-7.6+0.1\Delta_{1}-0.1\Delta_{3}-1.1\Delta_{5}+0.3\epsilon_{1})\cdot10^{-3}%
$\\
$y_{2}/\tilde{a}_{1}$ & $\epsilon_{2}t_{\beta}^{2}\lambda^{6}$ &
$(4.4-0.1\Delta_{1}+1.1\Delta_{4}+1.4\Delta_{5}-0.4\delta_{1}-0.4\epsilon_{1}%
$\\
&  & \quad$+3.0\epsilon_{2}-2.5\epsilon_{4}-4.8\delta_{5})\cdot10^{-4}%
$\\\hline
$x_{2}/\tilde{a}_{2}$ & $\delta_{2}$ & $-0.38-0.01\Delta_{1}-0.02\Delta
_{4}-0.01\delta_{1}+0.02\delta_{2}$\\\hline
$y_{3}/\tilde{a}_{3}$ & $\epsilon_{3}t_{\beta}^{2}\lambda^{6}$ &
$(-1.53-0.02\Delta_{1}+0.01\Delta_{1}+0.02\Delta_{3}+0.03\Delta_{4}%
-0.2\Delta_{5}$\\
&  & \quad$-0.02\delta_{1}+0.04\epsilon_{3}-0.2\delta_{5})\cdot10^{-2}$\\
$w_{1}/\tilde{a}_{3}$ & $0$ & $\sim-10^{-6}$\\\hline
$y_{4}/\tilde{a}_{4}$ & $c_{t}/y_{t}+\epsilon_{4}t_{\beta}^{2}\lambda^{6}$ &
$(-4.1-0.1\Delta_{4}-0.6\Delta_{5}+2.3\epsilon_{4})\cdot10^{-3}$\\
$w_{2}/\tilde{a}_{4}$ & $y_{c}/y_{t}+\eta_{2}\lambda^{4}$ & $(5.1-0.4\Delta
_{4}+0.7\eta_{2})\cdot10^{-3}$\\\hline
$y_{5}/\tilde{a}_{5}$ & $c_{b}/y_{b}+\delta_{5}$ & $-0.079-0.012\Delta
_{4}-0.003\Delta_{5}+0.084\delta_{5}$\\
$w_{3}/\tilde{a}_{5}$ & $y_{s}/y_{b}+\eta_{3}\lambda^{2}$ & \hskip 0.31 cm
$0.019-0.002\Delta_{5}+0.005\eta_{3}$\\
$w_{4}/\tilde{a}_{5}$ & $\eta_{4}\lambda^{4}$ & $(1.0+0.1\Delta_{4}%
-0.1\Delta_{5}+0.2\eta_{3}+2.3\eta_{4})\cdot10^{-4}$\\\hline
\end{tabular}
\caption{Dependence of the MFV parameters on the initial conditions when these
are taken in the neighbourhood of the SPS-1a benchmark point. Any correction
below one percent has been omitted.}%
\label{tab:sps1}%
\end{table}

The analytical solutions discussed above allow us to clearly understand the
origin of these fixed points. First of all, we stress that indeed the running
of the $\tilde{a}_{i}$'s is almost independent from the $x_{i}$'s, (which come
in only through the combination $S$), and is strongly dominated by the term
proportional to the gluino mass and strong gauge coupling, cf.
Eq.~(\ref{eq:a123}). Table~\ref{tab:sps1} shows that the $\tilde{a}_{i}$'s
tend to the value $8|M_{3}|^{2}/9$ (for $=1,2,3$) at the low scale, up to
small corrections, which of course can be calculated exactly with
Eq.~(\ref{eq:a123}). For $\tilde{a}_{4,5}$, we do not have an explicit
analytical solution, but from Eq.~(\ref{eq:mastersol}) we see that the
integral over the function $f$ tends to a value of the order $y_{t,b}M_{3}(t)$
at the low scale. For $\tilde{a}_{4}$ ($\tilde{a}_{5}$), the integral
$G_{\tilde{a}_{4(5)}}$ decreases (increases) from $1$ to about $0.4$
($3.3$) in going from the GUT to the electroweak scale, and so we
understand why the proportionality factor is smaller (larger) than one at
the electroweak scale. 

The analytical solutions for the $x_{i}$'s, Eqs.~(\ref{eq:n123sol},
\ref{eq:x12sol}), also explain the corresponding numbers in
Table~\ref{tab:sps1} and Fig.~\ref{fig:sps1}. First of all, the values of
$x_{1}$ and $x_{2}$ at the electroweak scale, with initial conditions
$x_{1}(t_{0})=x_{2}(t_{0})=0$, satisfy the relation $2x_{1}(t)=x_{2}(t)$
according to the solutions (\ref{eq:n123sol}, \ref{eq:x12sol}) -- a comparison
to the numerical solution of the exact RGE's shows that this is indeed
verified to better than one per mil\footnote{Notice that $x_{2}/a_{2}%
=2x_{1}/a_{2}=2x_{1}/a_{1}\cdot a_{1}/a_{2}=-0.36\cdot 1.06=-0.38$.}. We can
therefore consider the solution of only one of the two, which reads:
\begin{equation}
x_{1}(t)=\frac{m_{0}^{2}}{6}\left[  G_{n_{1}}(t)-1\right]  -\frac{G_{n_{1}%
}(t)}{3N}\!\!\int_{t}^{t_{0}}\!\!\!\!dt^{\prime}\frac{f_{n_{1}}(t^{\prime
})-f_{n_{2}}(t^{\prime})}{G_{n_{1}}(t^{\prime})}+\frac{1}{3N}\!\!\int
_{t}^{t_{0}}\!\!\!\!dt^{\prime}f_{n_{2}}(t^{\prime})\left[  1-\frac{G_{n_{1}%
}(t)}{G_{n_{1}}(t^{\prime})}\right]  . \label{eq:x1sol}%
\end{equation}
At the electroweak scale $t_{\text{ew}}$ the three terms contribute,
respectively
\begin{equation}
x_{1}(t_{\text{ew}})=-(1.3+73.1+2.9)\cdot10^{3}=-77.2\cdot10^{3}%
\;\mathrm{GeV}^{2}\;,
\end{equation}
which shows that the dominating contribution comes from the integral over
$f_{n_{1}}-f_{n_{2}}$ which contains only the $\tilde{a}_{i}$'s. In the ratio
$x_{1}/a_{1}$, this is the part which tends to a small constant value at the
electroweak scale, which almost looks like a fixed point. To understand why
this almost looks like a fixed point, we have to analyze the dependence of
$x_{1}(t)$ on the initial conditions. The relevant formulae have been given
above, cf. Eq.~(\ref{eq:x12ic}), and show a linear dependence, with
coefficients of order one. If we set, \emph{e.g.} $x_{1}(t_{0})=\tilde{a}_{1}%
(t_{0})\delta_{1}$ and $x_{2}(t_{0})=\tilde{a}_{2}(t_{0})\delta_{2}$, as
indicated in Table~\ref{tab:sps1}, then at the electroweak scale the effect is
equal to:
\begin{align}
\frac{\Delta x_{1}}{\tilde{a}_{1}}  &  =\delta_{1}\frac{\tilde{a}_{1}(t_{0}%
)}{\tilde{a}_{1}(t_{\text{ew}})}\frac{G_{n_{1}}(t_{\text{ew}})+5}{6}%
+\delta_{2}\frac{\tilde{a}_{2}(t_{0})}{\tilde{a}_{1}(t_{\text{ew}}%
)}\frac{G_{n_{1}}(t_{\text{ew}})-1}{6}\;,\\
\frac{\Delta x_{2}}{\tilde{a}_{2}}  &  =\delta_{1}\frac{\tilde{a}_{1}(t_{0}%
)}{\tilde{a}_{2}(t_{\text{ew}})}\frac{G_{n_{1}}(t_{\text{ew}})-1}{3}%
+\delta_{2}\frac{\tilde{a}_{2}(t_{0})}{\tilde{a}_{2}(t_{\text{ew}}%
)}\frac{G_{n_{1}}(t_{\text{ew}})+2}{3}\;,\nonumber
\end{align}
where $\Delta x_{i}\equiv x_{i}(t_{\text{ew}})_{|_{\delta_{k}\neq0}}%
-x_{i}(t_{\text{ew}})_{|_{\delta_{k}=0}}$, and inserting the numbers
$G_{n_{1}}(t_{\text{ew}})=0.228$, and $\tilde{a}_{1}(t_{0})/\tilde{a}_{1}%
(t_{\text{ew}})=0.023$, $\tilde{a}_{2}(t_{0})/\tilde{a}_{2}(t_{\text{ew}%
})=0.024$, valid for the initial conditions specified in Table~\ref{tab:sps1},
one perfectly reproduces the numbers there. Similarly one can explain why the
sensitivity to the initial conditions of the $\tilde{a}_{i}$'s is small: these
enter through the second integral in Eq.~(\ref{eq:x1sol}) and contribute as
follows. If we take $\tilde{a}_{i}(t_{0})=m_{0}^{2}(1+\Delta_{i})$ for
$i=1,2,3$ and $\tilde{a}_{4(5)}(t_{0})=y_{t(b)}(t_{0})A_{0}(1+\Delta_{4(5)})$
then we have (for $\Delta x_{1}\equiv x_{i}(t_{\text{ew}})_{|_{\Delta_{k}%
\neq0}}-x_{i}(t_{\text{ew}})_{|_{\Delta_{k}=0}}$)
\begin{align}
\Delta x_{1}  &  \!\!=\!\!-m_{0}^{2}(\Delta_{1}+\Delta_{2})\frac{G_{n_{1}}
(t_{\text{ew}})}{3N}\int_{t_{\text{ew}}}^{t_0}dt^{\prime}
\frac{6y_{t}^{2}(t^{\prime})}{G_{n_{1}}(t^{\prime})}
-y_{t}(t_{0})A_{0}\Delta_{4}\frac{G_{n_{1}}(t_{\text{ew}})}{3N}
\int_{t_{\text{ew}}}^{t_0}dt^{\prime}\frac{12\tilde{a}_{4}(t^{\prime})
G_{a_{4}}(t^{\prime}
)}{G_{n_{1}}(t^{\prime})}\nonumber\\
&  \!\!=\!\!-m_{0}^{2}\left[  0.13(\Delta_{1}+\Delta_{2})+0.38\Delta
_{4}\right]  \;.
\end{align}
Dividing these numbers by $\tilde{a}_{1}(t_{\text{ew}})$ one reproduces the
numbers in Table~\ref{tab:sps1}. Note that the sensitivity of $x_{1}%
/\tilde{a}_{1}$ to $\Delta_{1}$ receives another contribution due to the
denominator dependence on $\Delta_{1}$. This is easy to calculate and brings
the total coefficient below $1\%$ -- for this reason a term with $\Delta_{1}$
does not appear in the line for $x_{1}/\tilde{a}_{1}$. We stress again that
while the sensitivity of $x_{1}$ itself to the initial conditions of the
parameters which appear in its beta function is described by a coefficient
smaller than one, but not tiny, the ratio $x_{1}/\tilde{a}_{1}$ has tiny
coefficients because $\tilde{a}_{1}(t_{0})/\tilde{a}_{1}(t_{\text{ew}})$ is
tiny for the initial conditions given by the SPS-1a benchmark point.

All other coefficients appearing in Table~\ref{tab:sps1}, which specify the
sensitivity of the various parameters to the initial conditions of all the
others, can be understood analogously on the basis of our approximate
analytical formulae.

Finally, concerning the CP-violating phases, we see from Fig.~\ref{fig:sps1}
that if $\delta_{LR}^{U}$ or $\delta_{LR}^{D}$ have an imaginary part at the
GUT scale, even if large, it is much suppressed by the RGE effects. This is
easily understood on the basis of the simplified analytical solutions: the
imaginary part of $\tilde{a}_{4}$ misses the piece of the beta function which
is responsible for the fast running of the real part, cf.~(\ref{eq:a4a5}). At
the electroweak scale, only a residual phase remains, which acts as a small
correction to the CKM one present in $V_{ts}$ or $V_{td}$. Note that at
moderate $\tan\beta$, $\delta_{LR}^{D}$ is significantly affected by the
initial conditions for $\delta_{LR}^{U}$: a phase for the latter can generate
one for the former. On the other hand, any phase present in the trilinear
terms has only a very limited effect on $y_{2}$, relevant for $(\delta
_{LL})_{32}$ and $(\delta_{LL})_{31}$ mass-insertions. This follows also from
Table~\ref{tab:sps1}, where it is apparent that the initial conditions for the
trilinear terms affect only mildly the final value of $y_{2}$ at the
electroweak scale. Conversely, if $y_{2}\left(  t_{0}\right)  $ has a large
imaginary part, its running is very similar to the real part, i.e., it also
decreases dramatically (see upper right plot in Fig.~\ref{fig:sps1}). Further,
looking at the RGE's, $y_{2}$ cannot communicate its phase to the trilinear
terms. Therefore, all in all, if the SPS-1a is extended to include as many
CP-violating phases as allowed by MFV at the GUT scale, all these phases are
strongly suppressed at the electroweak scale. In that case, MFV essentially
collapses onto the real parameter case, usually considered in the literature
(see e.g. \cite{MFVpheno2}), where all CP-violating observables are tuned
entirely by the CKM phase.

\subsection{Running from the electroweak up to the GUT scale}

The numerical analysis of the SPS-1a benchmark point has shown that if one
assumes that the MFV hypothesis is valid at the GUT scale, one has rather
definite predictions at the low scale. We can now turn the question around and
ask how MFV evolves towards the high scale if one assumes to know the
parameters at the low scale. In particular, it is interesting to analyze what
happens if the boundary conditions at the low scale are chosen such that they
are rather far from the ``fixed point'' discussed above, but still well within
the naturalness assumption of MFV.

\begin{figure}[t]
\begin{center}
\includegraphics[width=9cm]{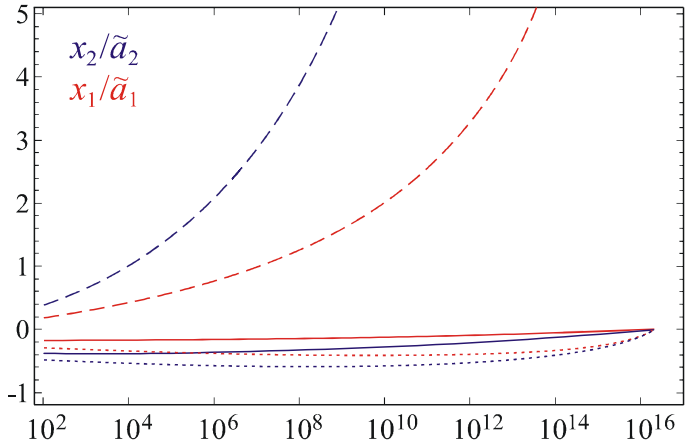}
\end{center}
\caption{RGE evolution of the MFV parameters for the SPS-1a benchmark point,
if the boundary conditions are fixed at the low scale. The dashed (dot-dashed)
lines correspond to the change of boundary conditions $x_{1(2)}(M_{Z})
=-x_{1(2)}^{\text{SPS-1a}}(M_{Z})$. The curves quitting the plot
reach $x_{1}/\tilde{a}_{1}\approx18$ and $x_{2}/\tilde{a}_{2}\approx40$ at the
GUT scale.}%
\label{fig:SPS1up}%
\end{figure}

We illustrate this in Fig.~\ref{fig:SPS1up}, in which we use as boundary
condition at $\mu=M_{Z}$ all the values obtained from the evolution
of the SPS-1a point, with the exception of $x_{1,2}$ for which we use
$x_{1,2}(M_{Z})=-x_{1,2}^{\text{SPS-1a}}(M_{Z})$.
The figure shows that a point far from the ``fixed point'' at the
low scale evolves to very high values of the MFV parameters at the high scale.
This behaviour is easily understood on the basis of our analytical solution:
from Table~\ref{tab:sps1} we read off that a change of natural size of the
initial conditions for $x_{1}$ produces a change of about $15\%$ at the low
scale. So, if we want to induce a change of $200\%$ at the low scale, we need
to make a change of more than one order of magnitude at the high scale. One
could imagine changing more parameters at the same time in such a way that the
evolution would be of MFV type all the way to the GUT scale.
Table~\ref{tab:sps1} shows, however, that this is not possible. Changing any
of the other parameter around the SPS-1a benchmark points, while staying in a
range compatible with MFV, does never generate a large change for
$x_{1}/\tilde{a}_{1}$ at the low scale (similar conclusions can be reached for
most other parameters).

The analysis of the previous section points to a possible way out: if the
ratio $\tilde{a}_{1}(t_{0})/\tilde{a}_{1}(t_{\text{ew}})$ were of order one
instead of $10^{-2}$, much larger changes in the low energy MFV parameters
would become possible. This ratio, however, is completely fixed by the RGE's
and initial conditions, cf. Eq.~(\ref{eq:a123}) and can be changed
significantly only by changing $\tilde{a}_{1}(t_{0})$ with respect to
$M_{3}(t_{0})$ such that the former becomes at least as large as the term
proportional to the gluino mass squared in Eq.~(\ref{eq:a123}). This, however,
brings us far away from the SPS-1a point, and moreover, will tend to make the
squarks heavier, which makes their contribution to low-energy flavour
violations, and correspondingly the phenomenological interest, smaller. Such a
situation is similar to the one realized in the case of the SPS-4 points
(independently of the size of $\tan\beta$), as we will discuss in the next section.

We conclude that, if the spectrum at the low scale corresponds to the SPS-1a
input at the high scale (note that the RGE's discussed here show that an
MFV-compatible change of the boundary conditions at the high scale has barely
any influence on the spectrum), and if the values of the MFV parameters
measured at the low scale are far away from those given in
Table~\ref{tab:sps1}, then the MFV hypothesis has to break down somewhere
before reaching the GUT scale.

This may indicate that either MFV has emerged from the RGE evolution of the
parameters, starting from a non MFV kind of MSSM, or that the MFV parameters
have a dynamical origin at a scale much lower than the GUT scale. The former
solution cannot be really investigated with our simplified RGE's, because if
any of the MFV parameters is far from its assumed order, our RGE's lose their
accuracy. One would have to study this solution with the full RGE's, and see
how much fine-tuned it is, or whether it can viewed as natural in any sense.
We leave this question open.

\section{Running MFV in the large $\tan\beta$ case}

\subsection{Analytical solutions}

When $\tan\beta\sim1/\lambda^{3}$, the parameters $y_{i}$ become in principle
of order one and cannot be neglected. The system of the RGE's is then not
amenable anymore to an analytical solution. The differential equations are all
coupled and a diagonalization, although possible in principle, is too
complicated to be of any use. Therefore, we start directly with the numerical
analysis and make some remarks about how one can understand
Table~\ref{tab:sps4} on the basis of the analytical solutions provided in the
previous section, plus some corrections specific to the large $\tan\beta$ case.

\subsection{Numerical example: the SPS-4 benchmark point}

\begin{figure}[pt]
\includegraphics[width=16cm]{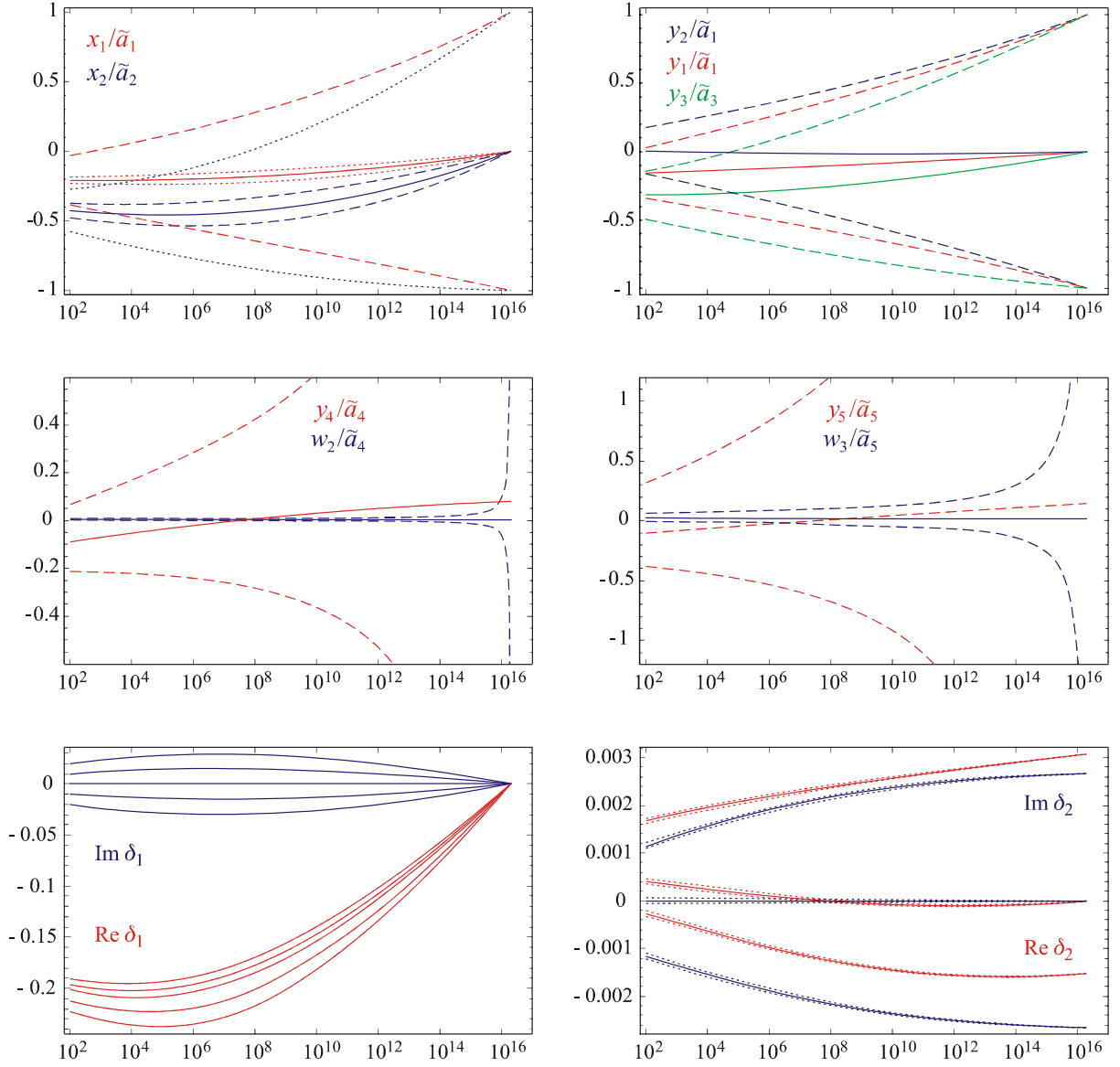}  \caption{Upper four plots: RGE
evolution of the MFV parameters for the SPS-4 benchmark point. The solid
curves always show the evolution of the mSUGRA type of initial conditions. In
the upper-left panel the dashed (dotted) lines show the evolution when
$x_{1(2)}(M_{\text{GUT}})=\pm M_{0}^{2}$. In all other cases, for each
parameter only three curves are shown -- the upper and lower ones (always
shown as dashed) correspond to different initial conditions for that single
parameter. Lower two plots: the mass-insertions $\delta_{1(2)}\equiv
(\delta_{RL}^{U(D)})^{32}/V_{ts}=(\delta_{RL}^{U(D)})^{31}/V_{td}$, with the
initial conditions at the GUT scales varied as explained in the text, but
allowing in addition for a large CP-phase (between $\pm180{{}^{\circ}}$). In
this case, $\delta_{1}\sim\tilde{a}_{4}$ is entirely radiatively generated,
since $\tilde{a}_{4}=0$ is set to zero at the GUT scale. The behaviours of the
other mass-insertions, Eqs.~(\ref{MIA1}, \ref{MIA2}, \ref{MIA3}), can easily
be obtained from those of the parameters shown in the upper four plots. In
particular, note that though $y_{2}\left(  M_{Z}\right)  $ is still smaller
than $x_{1}\left(  M_{Z}\right)  $, it is much less suppressed by RGE effects
than in the SPS-1a case, and so is its CP-violating phase.}%
\label{fig:sps4}%
\end{figure}

The SPS-4 benchmark point has $m_{0}=400$ GeV, $m_{1/2}=300$ GeV, $A_{0}=0$ at
the GUT scale, and $\tan\beta=50$. With these parameters, too large
supersymmetric contributions to $b\rightarrow s\gamma$ are generated, such
that the total rate comes out much lower than the measured one -- this point
is therefore not compatible with the phenomenology at the low scale. We use it
nonetheless for illustration purposes. Another important aspect is that
$A_{0}$ is set to zero at the GUT scale, and so $\tilde{a}_{4}=\tilde{a}_{5}%
=0$ there. If we want to vary the flavour violating parameters in the
trilinear couplings, we first need to define a scale of naturalness for them.
For this we use the value of $\tilde{a}_{4}$ and $\tilde{a}_{5}$ at the low
scale and set at $\mu=M_{\mathrm{GUT}}$
\begin{equation}
y_{4}\sim-y_{t}M_{3}t_{\beta}^{2}\lambda^{6}\;,\;\;\;\;w_{2}\sim-y_{t}%
M_{3}\lambda^{4}\;\;,
\end{equation}
and
\begin{equation}
y_{5}\sim-y_{b}M_{3}\;,\;\;\;\;w_{3}\sim-y_{b}M_{3}\lambda^{2}\;,\;\;\;\;w_{4}%
\sim-y_{b}M_{3}\lambda^{4}\;\;.
\end{equation}
As soon as we make any of these parameters different from zero, their ratio to
$\tilde{a}_{4}$ or $\tilde{a}_{5}$ at the high scale therefore diverges, as it
is seen in Fig.~\ref{fig:sps4}. Running down to the low scale, however, the
ratio quickly converges to its natural range, and becomes one or smaller.

The fact that $\tan\beta$ is so large means that most of the
$y_{i}$ and $w_{i}$ parameters change their order, since $\lambda^{3}\tan
\beta\sim0.6$. The curves in Figure~\ref{fig:sps4} show that the behaviour is
indeed different than for the SPS-1a benchmark point. Since now more
parameters are of order one (or almost), the mutual influences of the various
parameters are more important, though still quite small, as can be seen in
Table~\ref{tab:sps4}. The most prominent difference, however, is that the
``fixed points'' are now much weaker -- the value of the parameters at the low
scale is influenced more sensitively by the starting values at the high scale.
The same two statements can be made for the imaginary parts of the complex
coefficients: their suppression through RGE effects is now much weaker, and
the presence of a CP-phase for one parameter is felt more strongly by the
others. For example, if $\mathbf{A}^{U}$ and/or $\mathbf{A}^{D}$ involve
complex phases at the GUT scale, $y_{2}$ does develop a small but significant
phase at the low scale.

\begin{table}[t]%
\begin{tabular}
[c]{|l|l|l|}\hline
& $M_{\text{GUT}}$ & $M_{Z}$\\\hline
$\tilde{a}_{1}/M_{3}^{2}$ & $1.8(1+\Delta_{1})$ & $1.06+0.21\Delta
_{1}+0.01\Delta_{2}+0.006\Delta_{3}$\\
$\tilde{a}_{2}/M_{3}^{2}$ & $1.8(1+\Delta_{2})$ & $1.01+0.02\Delta
_{1}+0.17\Delta_{2}+0.02\Delta_{3}+0.008(\delta_{1}-2\delta_{2})$\\
$\tilde{a}_{3}/M_{3}^{2}$ & $1.8(1+\Delta_{3})$ & $1.00-0.01\Delta
_{1}+0.02\Delta_{2}+0.21\Delta_{3}-0.004(\delta_{1}-2\delta_{2})$\\
$\tilde{a}_{4}/(-y_{t}M_{3})$ & $\Delta_{4}$ & $0.78+0.09\Delta_{4}%
+0.02\Delta_{5}-0.09\epsilon_{4}-0.03\delta_{5}$\\
$\tilde{a}_{5}/(-y_{b}M_{3})$ & $\Delta_{5}$ & $0.93+0.02\Delta_{4}%
+0.16\Delta_{5}-0.03\epsilon_{4}-0.17\delta_{5}$\\\hline
$x_{1}/\tilde{a}_{1}$ & $\delta_{1}$ & $-0.21+0.01\Delta_{1}-0.02\Delta
_{2}-0.02\Delta_{4}+0.18\delta_{1}$\\
&  & $-0.02\delta_{2}+0.01\epsilon_{1}+0.02\epsilon_{4}-0.01\delta_{5}$\\
$y_{1}/\tilde{a}_{1}$ & $\epsilon_{1}t_{\beta}^{2}\lambda^{6}$ &
$-0.16+0.01\Delta_{1}-0.02\Delta_{3}-0.03\Delta_{5}+0.01\delta_{1}$\\
&  & $+0.07\epsilon_{1}-0.01\epsilon_{3}+0.02\delta_{5}$\\
$y_{2}/\tilde{a}_{1}$ & $\epsilon_{2}t_{\beta}^{2}\lambda^{6}$ &
$(0.6-0.1\Delta_{1}+0.6\Delta_{4}+0.7\Delta_{5}-1.4\delta_{1}-0.9\epsilon_{1}%
$\\
&  & \hskip 0.2 cm $+6.3\epsilon_{2}-0.9\epsilon_{4}-1.6\delta_{5}%
)\cdot10^{-2}$\\\hline
$x_{2}/\tilde{a}_{2}$ & $\delta_{2}$ & $-0.43-0.04\Delta_{1}+0.01\Delta
_{2}+0.01\Delta_{3}-0.03\Delta_{4}+0.01\Delta_{5}$\\
&  & \hskip 0.31 cm $-0.05\delta_{1}+0.15\delta_{2}-0.02\epsilon
_{1}-0.04\epsilon_{2}-0.01\epsilon_{4}+0.01\delta_{5}$\\\hline
$y_{3}/\tilde{a}_{3}$ & $\epsilon_{3}t_{\beta}^{2}\lambda^{6}$ &
$-0.32-0.046\Delta_{1}+0.012\Delta_{2}+0.02\Delta_{3}+0.01\Delta
_{4}-0.04\Delta_{5}$\\
&  & $-0.04\delta_{1}+0.01\delta_{2}-0.02\epsilon_{1}-0.03\epsilon
_{2}+0.06\epsilon_{3}-0.04\delta_{5}$\\
$w_{1}/\tilde{a}_{3}$ & $0$ & $\sim10^{-5}$\\\hline
$y_{4}/\tilde{a}_{4}$ & $\epsilon_{4}t_{\beta}^{2}\lambda^{6}$\quad$(^{\ast})$
& $-0.09-0.02\Delta_{4}-0.04\Delta_{5}+0.14\epsilon_{4}+0.02\delta_{5}$\\
$w_{2}/\tilde{a}_{4}$ & $\eta_{2}\lambda^{4}$ & $(5.0-1.1\Delta_{4}%
-0.1\Delta_{5}+0.3\epsilon_{4}+1.9\eta_{2}+0.3\delta_{5})\cdot10^{-3}$\\\hline
$y_{5}/\tilde{a}_{5}$ & $\delta_{5}$\hskip 1.15 cm $(^{\ast})$ &
$-0.10-0.046\Delta_{4}-0.023\Delta_{5}+0.015\epsilon_{4}+0.36\delta_{5}$\\
$w_{3}/\tilde{a}_{5}$ & $\eta_{3}\lambda^{2}$ & $0.023-0.005\Delta
_{5}+0.001\epsilon_{4}+0.003\delta_{5}+0.026\eta_{3}$\\
$w_{4}/\tilde{a}_{5}$ & $\eta_{4}\lambda^{4}$ & $(1.2+0.3\Delta_{4}%
-0.2\Delta_{5}+0.1\epsilon_{4}+0.2\delta_{5}+1.3\eta_{3}+7.8\eta_{4}%
)\cdot10^{-4}$\\\hline
\end{tabular}
\caption{Dependence of the MFV parameters on the initial conditions for the
SPS-4 point. Any correction below one percent has been omitted.\newline
$(^{\ast})$ Since $\tilde{a}_{4(5)}(M_{\text{GUT}})=0$, at this scale we
normalize $y_{4}$ and $w_{2}$ ($y_{5}$, $w_{3}$ and $w_{4}$) by $-y_{t}M_{3}$
($-y_{b}M_{3}$).}%
\label{tab:sps4}%
\end{table}

\begin{figure}[t]
\includegraphics[width=16cm]{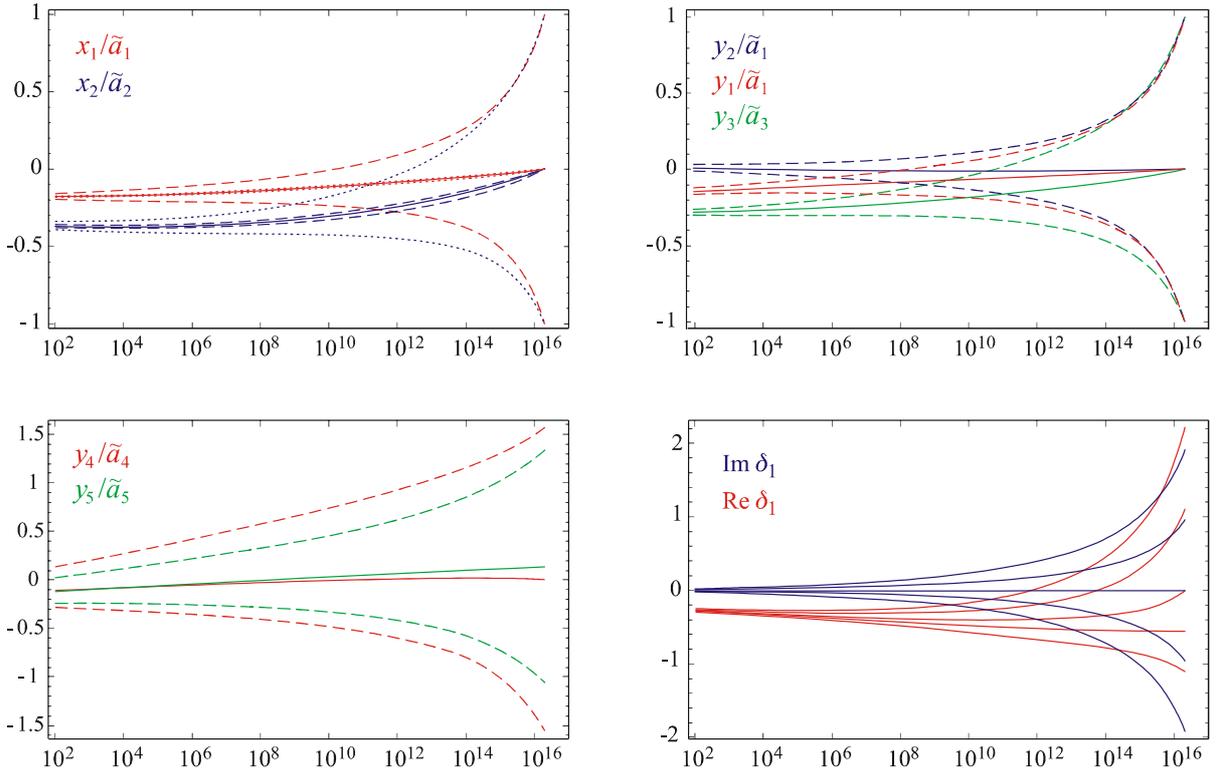}  \caption{RGE evolution of the MFV
parameters for the SPS-1a benchmark point, but with $\tan\beta=50$. As in
Fig.~\ref{fig:sps1}, the initial conditions at the GUT scale are varied, but
now allowing larger range for the $y_{i}$, which can be of $\mathcal{O}(1)$.
Still, the quasi fixed-point behaviour is obviously largely independent of
$\tan\beta$. The corresponding plot for $\delta_{2}$ is similarly close to the
$\tan\beta=10$ one. Those for $w_{1,2}$ are not shown because these parameters
stay very small, even with $\tan\beta=50$.}%
\label{fig:sps1a50}%
\end{figure}

Can we understand why this happens? Also, and more importantly, is it the
large $\tan\beta$ or the large $m_{0}$ which is responsible for the different
behaviour of the SPS-4 benchmark point? A first clue is given in
Fig.~\ref{fig:sps1a50}, where the boundary conditions are set according to the
SPS-1a, but for $\tan\beta=50$. Obviously, the fixed point behaviour is to a
large extent unaffected by $\tan\beta$, even for the $y_{i}$, which are now of
$\mathcal{O}(1)$. In particular, remark that $y_{2}$ still converge towards
zero, and so does its phase.

It is thus the larger $m_{0}$ which should be responsible for the different
behaviour shown in Fig.~\ref{fig:sps4}. This is confirmed with the help of the
analytical solutions, and Table~\ref{tab:sps4}. Consider for example
$x_{1}(t_{\text{ew}})$, which now depends ten times more strongly than in the
SPS-1a case on its initial condition $x_{1}(t_{0})$. If we ignore for a moment
that the beta function of $x_{1}$ also receives a contribution from $y_{2}$
and $y_{5}$, which are potentially large, and evaluate the coefficient in
front of $\delta_{1}$ in the seventh row of Table~\ref{tab:sps4} with the help
of Eq.~(\ref{eq:x12ic}) we get:
\begin{equation}
\frac{\Delta x_{1}(t_{\text{ew}})}{a_{1}(t_{\text{ew}})}=\delta_{1}%
\frac{a_{1}(t_{0})}{a_{1}(t_{\text{ew}})}\frac{G_{n_{1}}(t_{\text{ew}})+5}%
{6}=\delta_{1}\,0.21\cdot\frac{5.21}{6}=0.18\;,
\end{equation}
in perfect agreement with the number in the table. The reason for the enhanced
sensitivity to the initial conditions is due to the less strong growth of
$\tilde{a}_{1}$ in going from the GUT to the electroweak scale -- the extra
dependence on the initial conditions of $x_{1}$ coming from the terms
proportional to $y_{2}$ and $y_{5}^{2}$ is negligible, as we have explicitly
checked (the reason can be understood rather easily: the $y_{i}$'s depend on
$x_{1}$ through the beta function and vice-versa -- so the extra dependence of
$x_{1}(t_{\text{ew}})$ on its initial conditions at the GUT scale comes in
through the beta function of the beta function and is therefore suppressed by
two powers of $1/N$). In fact, even the value of $x_{1}(t_{\text{ew}})$ for
initial conditions of the mSUGRA type can be understood rather well on the
basis of the analytical formulae obtained in the case of moderate $\tan\beta$:
the relation $2x_{1}(t)=x_{2}(t)$ holds to better than one percent and
Eq.~(\ref{eq:x1sol}) yields numerically:
\begin{equation}
x_{1}(t_{\text{ew}})=-(2.10+13.76+0.42)\cdot10^{4}=-16.28\cdot10^{4}%
\;\mathrm{GeV}^{2}\;,
\end{equation}
whereas the value obtained solving the exact RGE's numerically is
$-16.11\cdot10^{4}$ GeV$^{2}$ -- the approximate analytical solutions obtained
in the moderate $\tan\beta$ case work here surprisingly well, to one
percent accuracy. 

\section{Conclusions}

In this paper, we have revisited the formulation of minimal flavour violation
(MFV) within Minimal Supersymmetric extensions of the Standard Model (MSSM),
and linked it to a counting rule which keeps track of the hierarchies in the
Yukawa couplings and in the CKM matrix in a coherent way. This allowed us to
move continuously and in a controlled manner, between the moderate and the
large $\tan\beta$ case, keeping control over the expected order of magnitude
of the different terms. We have argued that to implement these counting rules
in an efficient way, it is convenient to express the soft SUSY breaking terms
of the MSSM in a different basis than in the conventional MFV. In order to
study the renormalization group equations of the MFV parameters, we have
projected the beta functions of the soft SUSY breaking terms on this basis and
checked that the beta functions obey the same counting rules as the
coefficients themselves. We have then studied the behaviour of the running of
the MFV parameters numerically, and first checked that this reproduces the
full running with MFV boundary conditions at the level of accuracy at which we
expected them to work.

In the moderate $\tan\beta$ case, we were able to provide approximate
analytical solutions to the RGE's of the MFV parameters, and we discussed, in
the case of the SPS-1a benchmark point, the behaviour of the MFV parameters
under the running. We confirmed the finding of Ref.~\cite{Paradisi:2008qh}
about a quasi fixed-point behaviour of these, and could explain it in detail
with the help of our analytical solutions. The crucial observation is that the
flavour blind part of the soft SUSY-breaking terms, the $a_{i}$ parameters,
run fast, like the gluino masses, whereas all the terms with a nontrivial
flavour structure (the $x_{i}$'s, in our basis) grow less rapidly -- their
beta functions do not contain the gaugino masses, but at best the $a_{i}$'s.
This hierarchy in the beta functions explains why the ratios of the
flavour-violating parameters to the flavour-blind ones tend to a small, finite
value at the low scale. The only possibility we have identified to avoid this,
and to make a larger range of values at the low scale at all possible, is to
increase the value of the $a_{i}$'s at the GUT scale -- this, however, makes
the squark masses heavier, which in turn suppresses flavour violations at low energies.

Our analysis of the SPS-4 benchmark point, which has $\tan\beta=50$, shows
that the picture does not change substantially for large $\tan\beta$. More
parameters (the $y_{i}$'s in our notation) may become of order one, but their
RGE's are not particularly different from those of the $x_{i}$'s (they also
contain at best the $a_{i}$'s in their beta functions, and do not grow as
fast), nor do they influence the other parameters of order one in a
substantial way.

The picture which emerges from this analysis is that, if MFV is valid and has
its origin at the GUT scale (or at any other high scale much higher than the
SUSY breaking one), then it is a much more constraining framework than if one
assumes it to be valid just at the electroweak scale.

This is particularly true for the CP-violating phases. Throughout our study,
we have allowed MFV coefficients to have imaginary parts, whenever allowed by
the hermiticity of squark soft SUSY breaking mass terms. The first important
observation is that MFV does allow for new CP-violating phases not only in the
trilinear terms, but also in the LL sector. This latter phase could have
significant impacts on the MFV predictions for $b\rightarrow s$ and
$b\rightarrow d$ observables (but not for $s\rightarrow d$ ones, where it is
absent). However, if MFV has its origin at the GUT scale, and $\tan\beta$ is
not too large, we have shown that all the CP-violating phases are strongly
suppressed at the low-scale. This behaviour is essentially unaltered when
$\tan\beta$ is large, though in that case, the initial conditions at the GUT
scale have to be modified. In particular, when squark soft SUSY breaking mass
terms are as large as in the SPS-4 benchmark point, the suppression is much
less pronounced.

A final comment about leptons, which in this analysis have only played a
marginal role. One can of course apply similar ideas in that sector also, as
has been shown by Ref.~\cite{Cirigliano:2005ck}. Moreover, if one considers a
grand unified theory, one has a more constrained framework, and has relations
between the MFV parameters in the lepton and in the quark
sector~\cite{Grinstein:2006cg} (indeed, independently from any MFV hypothesis,
the relations between the two sectors have interesting phenomenological
implications, \cite{Ciuchini:2007ha}). While these connections are very
interesting and worth investigating, they mostly concern the boundary
conditions at the GUT scale -- the RGE below the GUT scale are the standard
MSSM ones. As a first analysis of the running of the MFV parameters, we
therefore found it more convenient to concentrate ourselves only on the quark sector.

\section*{Acknowledgments}

We thank Werner Porod for discussions and for making a beta version of his
program SPheno \cite{Porod:2003um} (which we have used for checks) available
to us prior to publication. This paper has been completed while two of us,
G.C. and C.S., were participating at the Workshop ``Flavour as a Window to New
Physics at the LHC''. We thank the organizers, Robert Fleischer, Thomas Mannel
and Yosef Nir for the invitation and Cern for hospitality. We acknowledge
stimulating discussions on the issues discussed here with several of the
participants. This work has been supported in part by the Swiss National
Foundation and by the EU, Contract No. MRTN-CT-2006-035482, ``FLAVIAnet''.\pagebreak 

\appendix

\section{Solving the RGE's and boundary conditions\label{AppendixRGE}}

For our purpose, only the one-loop RGE's are needed. At that order, the gauge,
Yukawa, soft-breaking and Higgs sectors essentially decouple, and solving the
RGE's proceeds in steps. First, the RGE's for the gauge couplings are solved,%
\begin{equation}
\frac{d\alpha_{i}\left(  t\right)  }{dt}=\frac{1}{2\pi}\beta_{i}\alpha_{i}%
^{2}\;\;,\;\;\;\beta_{i}=\left(  33/5,\,1,\,-3\right)  \;.
\end{equation}
Using, the $\overline{MS}$ values for simplicity, i.e. $\alpha_{em}%
^{-1}\left(  M_{Z}\right)  =127.904(17)$, $\sin^{2}\theta_{W}\left(
M_{Z}\right)  =0.23122(15)$, and $\alpha_{s}\left(  M_{Z}\right)
=0.1176(20)$, the unification scale and the corresponding value of the gauge
coupling are found as%
\begin{equation}
M_{G}=2.0\times10^{16}\;\text{GeV},\ \ \alpha_{G}^{-1}(M_{G})=24.3\;,
\end{equation}
if the MSSM running starts at the $M_{Z}$ scale, which we also assume for
simplicity. The gaugino masses are required to unify at the same scale $M_{G}$
as the couplings,%
\begin{equation}
M_{1}\left(  M_{G}\right)  =M_{2}\left(  M_{G}\right)  =M_{3}\left(
M_{G}\right)  \equiv m_{1/2}\;.
\end{equation}
Their RGE's are also solved to one-loop.

\begin{figure}[t]
\begin{center}
\includegraphics[width=16cm]{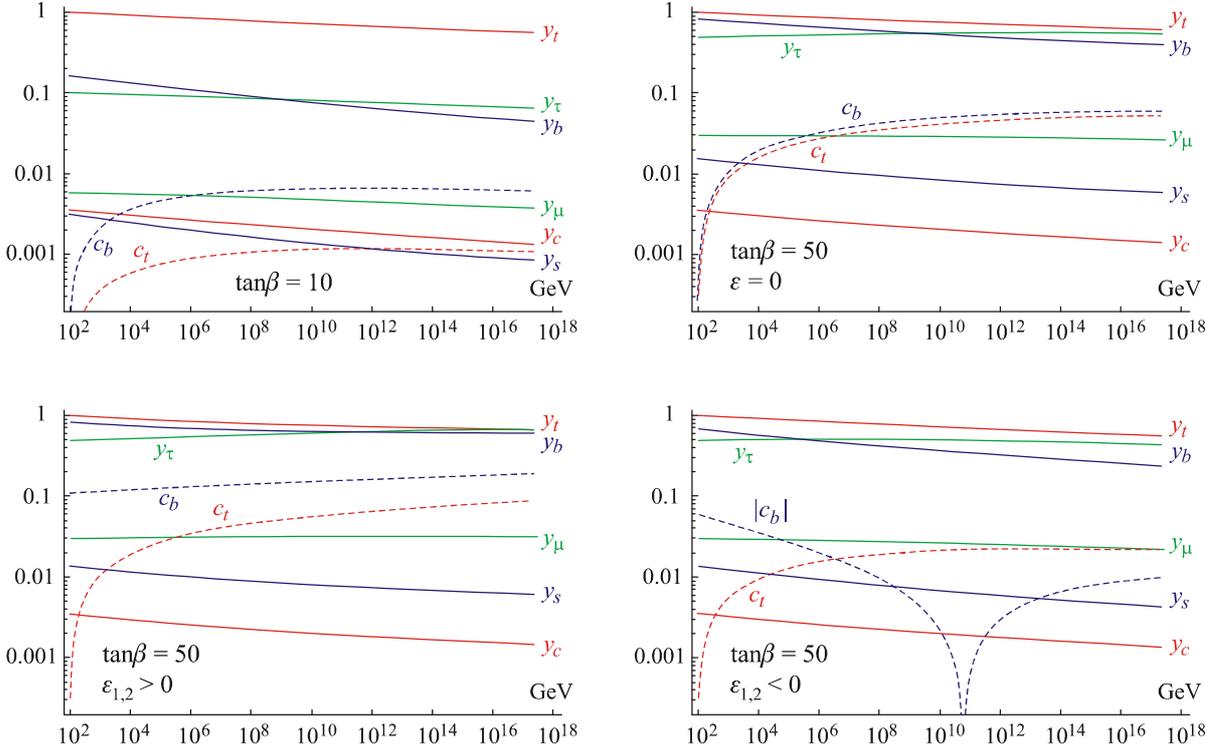}
\end{center}
\caption{Up: RGE evolution of the Yukawa couplings, in terms of their skeleton
structures given in Eq.(\ref{eq:Yusc}), for $\tan\beta=10$ and $50$. The
running obtained by solving the RGE of Eq.(\ref{eq:ysRGE}) or the full running
projected back on Eq.(\ref{eq:Yusc}) are indistinguishable. Down: the running
of the parameters with large $\tan\beta$ initial conditions, as specified in
Eq.~(\ref{EffLTanB}). In the right plot, starting with $c_{b}\left(
M_{Z}\right)  <0$, the RGE drives it positive, hence the zero at about
$10^{11}$ GeV.}%
\label{fig:YukawaRGE}%
\end{figure}

Second, the Yukawa couplings at the electroweak scale are set from the known
fermion masses at that scale (again, we neglect the difference between
$\overline{DR}$ and $\overline{MS}$ values for simplicity)
\begin{equation}%
\begin{array}
[c]{ccc}%
m_{u}\left(  M_{Z}\right)  =1.27\;\mathrm{MeV}\;, & m_{c}\left(  M_{Z}\right)
=0.619\;\mathrm{GeV}\;, & m_{t}\left(  M_{Z}\right)  =171.7\;\mathrm{GeV}\;,\\
m_{d}\left(  M_{Z}\right)  =2.9\;\mathrm{MeV}\;, & m_{s}\left(  M_{Z}\right)
=0.055\;\mathrm{GeV}\;, & m_{b}\left(  M_{Z}\right)  =2.89\;\mathrm{GeV}\;,\\
m_{e}\left(  M_{Z}\right)  =0.4866\;\mathrm{MeV}\;, & m_{\mu}\left(
M_{Z}\right)  =0.1027\;\mathrm{GeV}\;, & m_{\tau}\left(  M_{Z}\right)
=1.7462\;\mathrm{GeV}\;.
\end{array}
\end{equation}
as well as by setting $\tan\beta\equiv v_{u}/v_{d}$ and $v_{u}^{2}+v_{d}%
^{2}\simeq(174$\ $\mathrm{GeV})^{2}$. Their one-loop RGE's can then
immediately be solved numerically. Running up according to either the full
one-loop MSSM beta functions, or with the approximate MFV RGE's, we find for
$\tan\beta=10$:%
\begin{align}
|\mathbf{Y}_{u}^{\text{MFV}}\left(  M_{G}\right)  |  &  =\left(
\begin{array}
[c]{ccc}%
0 & 0 & 0\\
0.00032 & 0.0014 & 6.0\cdot10^{-5}\\
0.00459 & 0.0236 & 0.577
\end{array}
\right)  ,\;\;\Delta\mathbf{Y}_{u}\lesssim\left(
\begin{array}
[c]{ccc}%
10^{-6} & 10^{-6} & 10^{-8}\\
10^{-8} & 10^{-7} & 10^{-7}\\
10^{-6} & 10^{-6} & 10^{-4}%
\end{array}
\right)  \;,\nonumber\\
|\mathbf{Y}_{d}^{\text{MFV}}\left(  M_{G}\right)  |  &  =\left(
\begin{array}
[c]{ccc}%
0 & 0 & 0\\
0 & 0.000911 & 0\\
5.03\cdot10^{-5} & 0.000259 & 0.0545
\end{array}
\right)  ,\;\;\Delta\mathbf{Y}_{d}\lesssim\left(
\begin{array}
[c]{ccc}%
10^{-4} & 10^{-9} & 10^{-7}\\
10^{-8} & 10^{-7} & 10^{-6}\\
10^{-8} & 10^{-7} & 10^{-6}%
\end{array}
\right)  \;,\nonumber\\
\mathbf{Y}_{e}^{\text{MFV}}\left(  M_{G}\right)   &  =\mathrm{diag}\left(
0,\;0.0040,\;0.0675\right)  ,\;\;\Delta\mathbf{Y}_{e}\lesssim\mathrm{diag}%
\left(  10^{-5},\;10^{-8},\;10^{-7}\right)  \;, \label{App10}%
\end{align}
where $\Delta\mathbf{Y}=|\mathbf{Y}_{u}^{\text{MFV}}\left(  M_{G}\right)
|-|\mathbf{Y}_{u}^{\text{Full}}\left(  M_{G}\right)  |$, and for $\tan
\beta=50$ (neglecting non-holomorphic corrections):%
\begin{gather}
|\mathbf{Y}_{u}^{\text{MFV}}\left(  M_{G}\right)  |=\left(
\begin{array}
[c]{ccc}%
0 & 0 & 0\\
0.00034 & 0.00146 & 6.2\cdot10^{-5}\\
0.0050 & 0.0259 & 0.683
\end{array}
\right)  ,\;\;\Delta\mathbf{Y}_{u}\lesssim\left(
\begin{array}
[c]{ccc}%
10^{-6} & 10^{-6} & 10^{-8}\\
10^{-6} & 10^{-6} & 10^{-5}\\
10^{-4} & 10^{-4} & 10^{-2}%
\end{array}
\right)  \;,\nonumber\\
|\mathbf{Y}_{d}^{\text{MFV}}\left(  M_{G}\right)  |=\left(
\begin{array}
[c]{ccc}%
0 & 0 & 0\\
0 & 0.00613 & 0\\
0.00047 & 0.00242 & 0.472
\end{array}
\right)  ,\;\;\Delta\mathbf{Y}_{d}\lesssim\left(
\begin{array}
[c]{ccc}%
10^{-4} & 10^{-8} & 10^{-7}\\
10^{-7} & 10^{-5} & 10^{-5}\\
10^{-5} & 10^{-5} & 10^{-3}%
\end{array}
\right)  \;,\nonumber\\
\mathbf{Y}_{e}^{\text{MFV}}\left(  M_{G}\right)  =\mathrm{diag}\left(
0,\;0.0266,\;0.542\right)  ,\;\;\Delta\mathbf{Y}_{e}\lesssim\mathrm{diag}%
\left(  10^{-4},\;10^{-4},\;10^{-3}\right)  \;. \label{App50}%
\end{gather}
This shows that our counting is valid at all scales. It also shows in practice
how it is enforced: entries of absolute size at or below the $\mathcal{O}%
(10^{-4})$ are neglected (for consistency, since $m_{u}$, $m_{d}$ and $m_{e}$
are set to zero), while the non zero entries are reproduced up to
$\mathcal{O}(10^{-2})$ corrections. This structure is shared by the squark and
slepton soft-breaking terms, following Eq.~(\ref{eq:xMFV}). The evolution of
the parameters used to describe the running of the full Yukawa couplings is
shown in Fig.~\ref{fig:YukawaRGE}. The scalings of these parameters, in powers
of $\lambda$, can be immediately read off these plots. We stress, however,
that higher order terms in the RGE, as well as in the matching at the
electroweak scale, are relevant for the precise values of the Yukawa couplings
at the GUT scale. In particular, at the one-loop level, the Yukawa couplings
fail to unify at the GUT scale, as shown in Fig.~\ref{fig:YukawaRGE}. This is
of no concern for the present work. Rather, our purpose was to check whether
the beta functions expanded according to our counting rules do indeed
reproduce very precisely the full one loop RGE. As shown in Eqs.~(\ref{App10},
\ref{App50}), this is indeed the case, and there is a priori nothing
preventing one to extend the present analysis to include higher order effects.

In particular, to show the impact of the non-holomorphic corrections, we
consider the simulated scenario corresponding to%
\begin{equation}
\mathbf{\Delta}=\tan\beta\left(  \varepsilon_{1}\mathbf{Y}_{u}^{\dagger
}\mathbf{Y}_{u}+\varepsilon_{2}\mathbf{Y}_{d}^{\dagger}\mathbf{Y}_{d}\right)
\;, \label{EffLTanB}%
\end{equation}
with $\tan\beta=50$ and $\varepsilon_{1}=\pm0.002$, $\varepsilon_{2}=\pm
0.003$, which is in the ballpark of the values given e.g. in
Ref.~\cite{LargeTanb}, but still small enough that $\eta_{i}\approx1$ in
Eq.~(\ref{LTB:series}). Such a correction can be accounted for by shifting the
initial conditions for $y_{s}\left(  M_{Z}\right)  $, $y_{b}\left(
M_{Z}\right)  $ and $c_{b}\left(  M_{Z}\right)  $. Solving
Eq.~(\ref{LTB:background}) iteratively, we find
\begin{align}
\varepsilon_{1}  &  =+0.002,\varepsilon_{2}=+0.003\Rightarrow y_{s}\left(
M_{Z}\right)  =0.0138,\;y_{b}\left(  M_{Z}\right)  =0.832,\;c_{b}\left(
M_{Z}\right)  =0.110\;,\\
\varepsilon_{1}  &  =-0.002,\varepsilon_{2}=-0.003\Rightarrow y_{s}\left(
M_{Z}\right)  =0.0138,\;y_{b}\left(  M_{Z}\right)  =0.683,\;c_{b}\left(
M_{Z}\right)  =0.0595\;.
\end{align}
Taking these initial conditions, and running the Yukawa couplings according to
Eq.~(\ref{eq:ysRGE}), or using the full MSSM running, is numerically
equivalent to within one part in $10^{-4}$, as in Eqs.~(\ref{App10},
\ref{App50}). The behavior of the parameters is shown in
Fig.~\ref{fig:YukawaRGE}. As one can see, the parameter $c_{b}\left(
\mu\right)  $, though now slightly larger, is still sufficiently small to be
counted as of order $\mathcal{O}(y_{b}\lambda^{2})$.

The third step is to solve the RGE's for $m_{Hu}^{2}$, $m_{Hd}^{2}$,
$\mathbf{m}_{Q}^{2}$, $\mathbf{m}_{U}^{2}$, $\mathbf{m}_{D}^{2}$,
$\mathbf{A}^{U}$, $\mathbf{A}^{D}$, $\mathbf{m}_{L}^{2}$, $\mathbf{m}_{E}^{2}%
$, and $\mathbf{A}^{E}$, setting initial conditions at the high scale (details
of which are given in the text). Typically, we start with
\[
m_{Hu}^{2}\left(  M_{G}\right)  =m_{Hd}^{2}\left(  M_{G}\right)  \equiv
m_{0}^{2}\;,
\]
with $m_{0}\approx A_{0}$ also setting the scale of the squark and slepton
soft-breaking terms.

The final step is to enforce the matching at the electroweak scale for the
Higgs sector. The Higgs vacuum expectation values $v_{u}$ and $v_{d}$ are
fixed from $\tan\beta$ and $M_{Z}$. Enforcing the correct Higgs potential, and
knowing $m_{Hu}^{2}\left(  M_{Z}\right)  $ and $m_{Hd}^{2}\left(
M_{Z}\right)  $, gives $|\mu\left(  M_{Z}\right)  |$ and $b\left(
M_{Z}\right)  $, which can then be run up to the $M_{G}$ scale. Lowest order
approximations are notoriously inadequate at this step, but this is of no
concern for us since this matching and running is fully decoupled from the
rest, in particular from the squark and slepton sector on which we concentrate.

\section{Relation between the $x_{i}$ and the $b_{i}$
parameters\label{AppendixBasis}}

The standard MFV basis of Ref.~\cite{D'Ambrosio:2002ex}, and the one with the
$X_{i}$ matrices adopted here are related to each other by a linear
transformation. The MFV representation of the soft SUSY breaking terms given
in Eq.~(\ref{eq:xMFV}), corresponds to the following in the standard basis
(assuming real MFV coefficients for simplicity):
\begin{align}
\mathbf{m}_{Q}^{2}  &  =m_{0}^{2}\left[  a_{1}+b_{1}\mathbf{Y}_{u}^{\dagger
}\mathbf{Y}_{u}+b_{2}\mathbf{Y}_{d}^{\dagger}\mathbf{Y}_{d}+c_{1}%
(\mathbf{Y}_{d}^{\dagger}\mathbf{Y}_{d}\mathbf{Y}_{u}^{\dagger}\mathbf{Y}%
_{u}+\mathbf{Y}_{u}^{\dagger}\mathbf{Y}_{u}\mathbf{Y}_{d}^{\dagger}%
\mathbf{Y}_{d})\right]  \;,\nonumber\\
\mathbf{m}_{U}^{2}  &  =m_{0}^{2}\left[  a_{2}+b_{3}\mathbf{Y}_{u}%
\mathbf{Y}_{u}^{\dagger}\right]  \;,\nonumber\\
\mathbf{m}_{D}^{2}  &  =m_{0}^{2}\left[  a_{3}+\mathbf{Y}_{d}\left(
b_{6}+b_{7}\mathbf{Y}_{u}^{\dagger}\mathbf{Y}_{u}\right)  \mathbf{Y}%
_{d}^{\dagger}\right]  \;,\nonumber\\
\mathbf{A}^{U}  &  =A_{0}\mathbf{Y}_{u}\,\left[  a_{4}+b_{9}\mathbf{Y}%
_{u}^{\dagger}\mathbf{Y}_{u}+b_{10}\mathbf{Y}_{d}^{\dagger}\mathbf{Y}%
_{d}\right]  \;,\nonumber\\
\mathbf{A}^{D}  &  =A_{0}\mathbf{Y}_{d}\,\left[  a_{5}+b_{11}\mathbf{Y}%
_{u}^{\dagger}\mathbf{Y}_{u}+b_{12}\mathbf{Y}_{d}^{\dagger}\mathbf{Y}%
_{d}+c_{6}\mathbf{Y}_{d}^{\dagger}\mathbf{Y}_{d}\mathbf{Y}_{u}^{\dagger
}\mathbf{Y}_{u}\right]  \;. \label{eq:rMFV}%
\end{align}

The connection between both bases is explicitly given here:%
\begin{align}
\left(
\begin{array}
[c]{c}%
\tilde{a}_{1}\\
x_{1}\\
y_{1}\\
y_{2}%
\end{array}
\right)   &  =m_{0}^{2}\left(
\begin{array}
[c]{cccc}%
1 & 0 & 0 & 0\\
0 & y_{t}^{2} & c_{b}^{2} & 2c_{b}y_{t}\bar{y}_{b}\bar{y}_{t}\\
0 & c_{t}^{2} & y_{b}^{2} & 2c_{t}y_{b}\bar{y}_{b}\bar{y}_{t}\\
0 & c_{t}y_{t} & c_{b}y_{b} & 2\left(  c_{t}c_{b}+y_{b}y_{t}\right)  \bar
{y}_{b}\bar{y}_{t}%
\end{array}
\right)  \left(
\begin{array}
[c]{c}%
a_{1}\\
b_{1}\\
b_{2}\\
c_{1}%
\end{array}
\right)  \;,\nonumber\\
\left(
\begin{array}
[c]{c}%
\tilde{a}_{2}\\
x_{2}%
\end{array}
\right)   &  =m_{0}^{2}\left(
\begin{array}
[c]{cc}%
1 & 0\\
0 & \bar{y}_{t}^{2}%
\end{array}
\right)  \left(
\begin{array}
[c]{c}%
a_{2}\\
b_{3}%
\end{array}
\right)  \;,\nonumber\\
\left(
\begin{array}
[c]{c}%
\tilde{a}_{3}\\
y_{3}\\
w_{1}%
\end{array}
\right)   &  =m_{0}^{2}\left(
\begin{array}
[c]{ccc}%
1 & 0 & 0\\
0 & \bar{y}_{b}^{2} & \bar{y}_{t}^{2}\bar{y}_{b}^{2}\\
0 & -A\lambda^{2}c_{b}y_{s} & -A\lambda^{2}y_{s}y_{t}\bar{y}_{b}\bar{y}_{t}%
\end{array}
\right)  \left(
\begin{array}
[c]{c}%
a_{1}\\
b_{6}\\
b_{7}%
\end{array}
\right)  \;,\nonumber\\
\left(
\begin{array}
[c]{c}%
\tilde{a}_{4}\\
y_{4}\\
w_{2}%
\end{array}
\right)   &  =A_{0}\left(
\begin{array}
[c]{ccc}%
y_{t} & y_{t}\bar{y}_{t}^{2} & c_{b}\bar{y}_{b}\bar{y}_{t}\\
c_{t} & c_{t}\bar{y}_{t}^{2} & y_{b}\bar{y}_{b}\bar{y}_{t}\\
y_{c} & 0 & 0
\end{array}
\right)  \left(
\begin{array}
[c]{c}%
a_{4}\\
b_{9}\\
b_{10}%
\end{array}
\right)  \;,\nonumber\\
\left(
\begin{array}
[c]{c}%
\tilde{a}_{5}\\
y_{5}\\
w_{3}\\
w_{4}%
\end{array}
\right)   &  =A_{0}\left(
\begin{array}
[c]{cccc}%
y_{b} & c_{t}\bar{y}_{t}\bar{y}_{b} & y_{b}\bar{y}_{b}^{2} & c_{t}\bar{y}%
_{t}\bar{y}_{b}^{3}\\
c_{b} & y_{t}\bar{y}_{t}\bar{y}_{b} & c_{b}\bar{y}_{b}^{2} & y_{t}\bar{y}%
_{t}\bar{y}_{b}^{3}\\
y_{s} & 0 & 0 & 0\\
0 & -A\lambda^{2}y_{s}y_{t}\bar{y}_{t} & -A\lambda^{2}y_{s}c_{b}\bar{y}_{b} &
-A\lambda^{2}y_{s}c_{b}\bar{y}_{b}\bar{y}_{t}^{2}%
\end{array}
\right)  \left(
\begin{array}
[c]{c}%
a_{5}\\
b_{11}\\
b_{12}\\
c_{6}%
\end{array}
\right)  \;,\nonumber\\
\left(
\begin{array}
[c]{c}%
\tilde{a}_{6}\\
y_{6}\\
\tilde{a}_{7}\\
y_{7}%
\end{array}
\right)   &  =m_{0}^{2}\left(
\begin{array}
[c]{cccc}%
1 & 0 & 0 & 0\\
0 & y_{\tau}^{2} & 0 & 0\\
0 & 0 & 1 & 0\\
0 & 0 & 0 & y_{\tau}^{2}%
\end{array}
\right)  \left(
\begin{array}
[c]{c}%
a_{6}\\
b_{13}\\
a_{7}\\
b_{14}%
\end{array}
\right)  \;,\nonumber\\
\left(
\begin{array}
[c]{c}%
\tilde{a}_{8}\\
w_{5}%
\end{array}
\right)   &  =A_{0}\left(
\begin{array}
[c]{cc}%
y_{\tau} & y_{\tau}^{3}\\
y_{\mu} & 0
\end{array}
\right)  \left(
\begin{array}
[c]{c}%
a_{8}\\
b_{15}%
\end{array}
\right)  \;.
\end{align}
From this, and the scaling of the Yukawa couplings, one gets the scaling of
the MFV coefficients, assuming all $a$'s and $b$'s are a priori of
$\mathcal{O}\left(  1\right)  $. The extension to complex MFV coefficients is immediate.

Note that the trilinear couplings have a non-trivial flavour structure even in
the limit $b_{i}=0$, and universality is reproduced by setting%
\begin{equation}%
\begin{array}
[c]{ccc}
& \tan\beta=10 & \tan\beta=50\dfrac{{}}{{}}\\
\left(
\begin{array}
[c]{c}%
\tilde{a}_{4}\\
y_{4}\\
w_{2}%
\end{array}
\right)  =a_{4}A_{0}\left(
\begin{array}
[c]{c}%
y_{t}\left(  M_{G}\right) \\
c_{t}\left(  M_{G}\right) \\
y_{c}\left(  M_{G}\right)
\end{array}
\right)  ,\; & a_{4}A_{0}\left(
\begin{array}
[c]{c}%
0.5762\\
0.0011\\
0.0014
\end{array}
\right)  ,\; & a_{4}A_{0}\left(
\begin{array}
[c]{c}%
0.6393\\
0.0052\\
0.0015
\end{array}
\right)  \;,\\
\left(
\begin{array}
[c]{c}%
\tilde{a}_{5}\\
y_{5}\\
w_{3}\\
w_{4}%
\end{array}
\right)  =a_{5}A_{0}\left(
\begin{array}
[c]{c}%
y_{b}\left(  M_{G}\right) \\
c_{b}\left(  M_{G}\right) \\
y_{s}\left(  M_{G}\right) \\
0
\end{array}
\right)  ,\; & a_{5}A_{0}\left(
\begin{array}
[c]{c}%
0.0481\\
0.0063\\
0.0009\\
0
\end{array}
\right)  ,\; & a_{5}A_{0}\left(
\begin{array}
[c]{c}%
0.4163\\
0.0589\\
0.0061\\
0
\end{array}
\right)  \;,\\
\left(
\begin{array}
[c]{c}%
\tilde{a}_{8}\\
w_{5}%
\end{array}
\right)  =a_{8}A_{0}\left(
\begin{array}
[c]{c}%
y_{\tau}\left(  M_{G}\right) \\
y_{\mu}\left(  M_{G}\right)
\end{array}
\right)  ,\; & a_{8}A_{0}\left(
\begin{array}
[c]{c}%
0.0675\\
0.0040
\end{array}
\right)  ,\; & a_{8}A_{0}\left(
\begin{array}
[c]{c}%
0.5432\\
0.0266
\end{array}
\right)  \;.
\end{array}
\end{equation}

\section{Higher order terms in the beta functions}

\label{app:la2}Following the counting rules given in Eqs.~(\ref{eq:counting},
\ref{xilambda}, \ref{xilambda2}, \ref{cbt}), the beta functions are expanded
in powers of $\lambda$. Actually, this expansion involves only even powers of
$\lambda$, so the first corrections arise at $\mathcal{O}(\lambda^{2})$ when
$\tan\beta\sim\lambda^{-3}$. For Yukawa couplings, slepton soft-breaking
terms, as well as for the Higgs parameters, the beta functions given in the
text are already precise to $\mathcal{O}(\lambda^{4})$ or higher, so only
those for squark soft-breaking terms are missing. Denoting $\delta\beta_{C}$
the $\mathcal{O}(\lambda^{2})$ corrections, they are easily found to be%
\begin{equation}%
\begin{array}
[c]{l}%
\delta\beta_{\tilde{a}1}=\delta\beta_{\tilde{a}2}=\delta\beta_{\tilde{a}3}%
=0\;,\smallskip\\
\delta\beta_{x_{1}}=2\left(  c_{b}y_{b}+c_{t}y_{t}\right)  \left(
y_{1}+\operatorname{Re}y_{2}\right)  \;,\smallskip\\
\delta\beta_{y_{1}}=2\left(  c_{b}y_{b}+c_{t}y_{t}\right)  \left(
y_{1}+\operatorname{Re}y_{2}\right)  \;,\smallskip\\
\delta\beta_{y_{2}}=\left(  c_{t}y_{t}+c_{b}y_{b}\right)  \left(
2\tilde{a}_{1}+x_{1}+y_{1}+2y_{2}\right)  +2c_{t}y_{t}\left(  m_{H_{u}}%
^{2}+\tilde{a}_{2}+x_{2}\right)  +2c_{b}y_{b}\left(  m_{H_{d}}^{2}%
+\tilde{a}_{3}+y_{3}\right)  \;,\medskip\\
\delta\beta_{x_{2}}=8c_{t}y_{t}\left(  m_{Hu}^{2}+\tilde{a}_{1}+\tilde{a}_{2}%
+x_{1}+x_{2}+y_{1}+2\operatorname{Re}y_{2}\right)  \smallskip\\
\delta\beta_{y_{3}}=8c_{b}y_{b}\left(  m_{Hd}^{2}+\tilde{a}_{1}+\tilde{a}_{3}%
+x_{1}+y_{1}+2\operatorname{Re}y_{2}+y_{3}\right)  \;,\smallskip\\
\delta\beta_{w_{1}}=-2A\lambda^{2}c_{b}y_{s}\left(  2m_{Hd}^{2}+2a_{1}%
+2a_{3}+2x_{1}+2y_{2}+y_{3}\right)  -2A\lambda^{4}\left(  \left(  x_{1}%
+y_{2}\right)  y_{b}y_{s}+w_{3}^{\ast}y_{5}\right)  \;,\medskip\\
\delta\beta_{\tilde{a}_{4}}=a_{4}\left(  25c_{t}y_{t}+c_{b}y_{b}\right)
+y_{4}\left(  11c_{t}y_{t}+c_{b}y_{b}\right)  +2y_{5}\left(  c_{t}y_{b}%
+c_{b}y_{t}\right)  \;,\smallskip\\
\delta\beta_{y_{4}}=a_{4}\left(  11c_{t}y_{t}+c_{b}y_{b}\right)  +y_{4}\left(
25c_{t}y_{t}+c_{b}y_{b}\right)  +2a_{5}\left(  c_{t}y_{b}+c_{b}y_{t}\right)
+c_{t}K_{u}^{\prime}\;,\smallskip\\
\delta\beta_{w_{2}}=6c_{t}\left(  \left(  a_{4}+y_{4}\right)  y_{c}+w_{2}%
y_{t}\right)  \;,\medskip\\
\delta\beta_{\tilde{a}_{5}}=a_{5}\left(  25c_{b}y_{b}+c_{t}y_{t}\right)
+y_{5}\left(  11c_{b}y_{b}+c_{t}y_{t}\right)  +2y_{4}\left(  c_{t}y_{b}%
+c_{b}y_{t}\right)  \;,\smallskip\\
\delta\beta_{y_{5}}=a_{5}\left(  11c_{b}y_{b}+c_{t}y_{t}\right)  +y_{5}\left(
25c_{b}y_{b}+c_{t}y_{t}\right)  +2a_{4}\left(  c_{t}y_{b}+c_{b}y_{t}\right)
+c_{b}\left(  2a_{8}y_{\tau}+K_{u}^{\prime}\right)  \;,\smallskip\\
\delta\beta_{w_{3}}=6c_{b}\left(  \left(  a_{5}+y_{5}\right)  y_{s}+w_{3}%
y_{b}\right)  \;,\smallskip\\
\delta\beta_{w_{4}}=2w_{4}\left(  5c_{b}y_{b}+c_{t}y_{t}\right)  -A\lambda
^{2}\left(  \left(  5c_{b}y_{b}+c_{t}y_{t}\right)  w_{3}+4c_{b}y_{s}\left(
a_{5}+y_{5}\right)  \right)  \smallskip\\
\;\;\;\;\;\;\;\;\;\;\;\;+\dfrac{A\lambda^{4}}{2}\left(  1-2\rho+2i\eta\right)
y_{t}\left(  w_{3}y_{t}+2\left(  a_{4}+y_{4}\right)  y_{s}\right)  \;.
\end{array}
\end{equation}
Including these corrections, the beta functions are all correct up to tiny
$\mathcal{O}(\lambda^{4})$ corrections. Numerically, they reproduce the full
MSSM one-loop running, projected back on the $X_{i}$ basis, to better than
$1\%$, and even to better than $0.1\%$ for mSUGRA-type of initial conditions
at the GUT scale. For the numerical analysis discussed in the text, we have
always used the beta functions including the corrections given here -- we
stress that even if formally suppressed, the terms containing $K_{u}^{\prime}$
or $m_{H_{u,d}}^{2}$ may be numerically important, whereas most of the others
give very small contributions.

\section{Fixed points}

\label{sec:fixedp} The numerical analysis has shown that all the MFV
parameters tend to a certain ``fixed point'' at the low scale -- rather
strongly for the SPS-1a point and less so for the SPS-4 one. Here, we discuss
this feature in more detail, and try to gain some analytical understanding of
it on the basis of the simple beta functions derived in Sect.~\ref{sec:RGE}. A
true fixed point occurs if the beta function of a parameter has a zero which
depends only on the value of that parameter -- depending on the sign of the
derivative of the beta function with respect to the parameter at the position
of the zero, the fixed point is an ultraviolet (positive) or an infrared one
(negative). In our case, we observe the behaviour of a quasi infrared fixed point.

Such behaviour is observed for the ratios $x_{1}/\tilde{a}_{1}$ and
$x_{2}/\tilde{a}_{2}$, for example, but also for all other similar ratios. The
beta functions for these ratios read:
\begin{equation}
\beta_{x_{i}/\tilde{a}_{i}}=\frac{1}{\tilde{a}_{i}}\left[  \beta_{x_{i}%
}-\frac{x_{i}}{\tilde{a}_{i}}\beta_{\tilde{a}_{i}}\right]  \;\;,
\end{equation}
and we have to find out whether these beta functions have zeros, and on what
parameters these zeros depend. The equations $\beta_{x_{i}/\tilde{a}_{i}}=0$
are unfortunately nonlinear, and all coupled to each other, such that an
analytical solution is difficult to obtain in general, and in any case, not
very illuminating because too cumbersome. The terms which make the equations
nonlinear, however, all originate from the dependence of the $\tilde{a}_{i}$'s
parameters on the $x_{i}$'s, etc., which, as we have seen, is negligible. The
behaviour of the $\tilde{a}_{i}$'s is mostly driven by the gluino masses, and
somewhat also by the terms proportional to $y_{t}$. In this approximation, it
is then easy to solve the equations and obtain simple analytical expressions
for the solutions. In addition, if one considers only moderate $\tan\beta$ and
neglects higher orders in $\lambda$, the equations simplify further, and look
as follows:
\begin{equation}
\frac{\bar{x}_{1}}{\tilde{a}_{1}}=-\frac{m_{H_{u}}^{2}+\tilde{a}_{1}%
+\tilde{a}_{2}+\tilde{a}_{4}/y_{t}^{2}}{\tilde{a}_{1}+2\tilde{a}_{2}%
+\frac{16}{3}g_{3}^{2}M_{3}^{2}/y_{t}^{2}}\qquad,\;\qquad\frac{\bar{x}_{2}%
}{\tilde{a}_{2}}=2\frac{\bar{x}_{1}}{\tilde{a}_{1}}\;.
\end{equation}
Analogous results can be given for all the other parameters. These zeros of
the beta functions are not true fixed points, because they depend on
parameters which do run -- they are rather some sort of ``running fixed
points''. This means that even if a parameter reaches exactly its ''fixed
point'' a certain scale, it will not stay there because the zero of the beta
function will move with the scale. On the other hand, even if moving with the
scale, they do represent a line of attraction for the parameters.

For the ratios $x_{1}/\tilde{a}_{1}$ and $x_{2}/\tilde{a}_{2}$, the running
fixed points flatten out at the low scale, such that in that region they
become almost true fixed points. Indeed, also the numerical evaluation shows
that they almost coincide with the values these parameters tend to (for the
SPS-1a point):
\begin{align}
\frac{\bar{x}_{1}}{\tilde{a}_{1}}_{|_{M_Z}}  &
=-0.18\;\;,\;\;\;\;\;\;\frac{x_{1}}{\tilde{a}_{1}}_{|_{M_Z}%
}=-0.18\;\;,\nonumber\\
\frac{\bar{x}_{2}}{\tilde{a}_{2}}_{|_{M_Z}}  &
=-0.39\;\;,\;\;\;\;\;\;\frac{x_{2}}{\tilde{a}_{2}}_{|_{M_Z}}=-0.38\;\;.
\end{align}
These formulae do not allow one to understand how strong these quasi fixed
points are at the low scale -- the proper explanation of the behaviour of the
RGE has been provided in Sect.~\ref{sec:SPS1a}

\end{document}